\documentclass{natureprintstyle}
\usepackage{amssymb,amsmath,graphicx}

\def\simlt{\mathrel{\hbox{\rlap{\hbox{\lower4pt\hbox{$\sim$}}}\hbox{$<$}}}}
\def\simgt{\mathrel{\hbox{\rlap{\hbox{\lower4pt\hbox{$\sim$}}}\hbox{$>$}}}}

\def\ale{\mathrel{\hbox{\rlap{\hbox{\lower4pt\hbox{$\sim$}}}\hbox{$<$}}}}
\def\age{\mathrel{\hbox{\rlap{\hbox{\lower4pt\hbox{$\sim$}}}\hbox{$>$}}}}

\newcommand{\cgs}{\mbox{${\rm erg~cm}^{-2}~\rm{s}^{-1}$}}
\newcommand{\batrate}{\mbox{${\rm cts~cm}^{-2}~\rm{s}^{-1}$}}

\newcommand{\meszaros}{M\'{e}sz\'{a}ros}

\newcommand{\swift}{{\it Swift}}

\newcommand{\transient}{{Sw~J1644+57}}

\begin{document}

\title{Discovery of the Onset of Rapid Accretion by a Dormant Massive Black Hole.}

\author{D.~N.~Burrows$^1$, 
J.~A.~Kennea$^1$,  
G.~Ghisellini$^2$, 
V.~Mangano$^3$,  
B.~Zhang$^4$, 
K.~L.~Page$^5$, 
M.~Eracleous$^1$,  
P.~Romano$^3$, 
T.~Sakamoto$^{6,7}$,   
A.~D.~Falcone$^1$, 
J.~P.~Osborne$^5$, 
S.~Campana$^2$, 
A.~P.~Beardmore$^5$,
A.~A.~Breeveld$^{8}$,
M.~M.~Chester$^1$,
R.~Corbet$^{6,7}$, 
S.~Covino$^2$,
J.~R.~Cummings$^{6,7}$, 
P.~D'Avanzo$^2$, 
V.~D'Elia$^{9}$, 
P.~Esposito$^{10}$,
P.~A.~Evans$^5$, 
D.~Fugazza$^2$,
J.~M.~Gelbord$^1$, 
K.~Hiroi$^{11}$, 
S.~T.~Holland$^{6,12}$, 
K.~Y.~Huang$^{13}$, 
M.~Im$^{14}$, 
G.~Israel$^{15}$,
Y.~Jeon$^{14}$, 
Y.-B.~Jeon$^{16}$,
N.~Kawai$^{17,20}$,
H.~A.~Krimm$^{6,12}$, 
P.~\meszaros$^1$,
H.~Negoro$^{18}$,
N.~Omodei$^{19}$,  
W.-K.~Park$^{14}$, 
J.~S.~Perkins$^{6,7}$, 
M.~Sugizaki$^{20}$,
H.-I.~Sung$^{16}$, 
G.~Tagliaferri$^2$,
E.~Troja$^{21,22}$,
Y.~Ueda$^{11}$,
Y.~Urata$^{23}$,
R.~Usui$^{17}$,
L.~A.~Antonelli$^{15,9}$, 
S.~D.~Barthelmy$^{21}$, 
G.~Cusumano$^3$, 
P.~Giommi$^{9}$, 
F. E. Marshall$^{21}$,
 A.~Melandri$^2$, 
M. Perri$^{9}$, 
J.~L.~Racusin$^{21}$, 
B. Sbarufatti$^3$, 
M.~H.~Siegel$^1$, 
\& N.~Gehrels$^{21}$}

\date{\today}{}
\maketitle

\begin{affiliations}
\item Department of Astronomy \& Astrophysics, The Pennsylvania
    State University, 525 Davey Laboratory, University Park, PA 16802,
    USA   %
\item INAF -- Osservatorio Astronomico di Brera, Via Bianchi 46,
    23807 Merate, Italy %
\item INAF -- Istituto di Astrofisica Spaziale e Fisica Cosmica, Via
    U.\ La Malfa 153, I-90146 Palermo, Italy %
\item Department of Physics and Astronomy, University of Nevada, Las
  Vegas, NV 89154, USA %
\item Department of Physics and Astronomy, University of Leicester, University
    Road, LE1 7RH Leicester, UK %
\item CRESST and NASA/GSFC, Greenbelt, MD 20771, USA. %
\item University of Maryland, Baltimore
     County, 1000 Hilltop Circle, Baltimore, MD 21250, USA %
\item University College London / Mullard Space Science Laboratory   %
\item ASI Science Data Center, via Galileo Galilei, 00044 Frascati, Italy %
\item INAF -- Osservatorio Astronomico di Cagliari, localit\`a
    Poggio dei Pini, strada 54, I-09012 Capoterra, Italy %
\item Department of Astronomy, Kyoto University, Oiwake-cho, Sakyo-ku,
  Kyoto 606-8502, Japan %
\item Universities Space Research Association, 10211 Wincopin Circle, Suite 500, 
Columbia, MD 21044-3432, USA  %
\item Academia Sinica Institute of Astronomy and Astrophysics,
    Taipei 106, Taiwan %
\item Center for the Exploration of the Origin of the Universe,
    Department of Physics \& Astronomy, FPRD, Seoul National
    University, Shillim-dong, San 56-1, Kwanak-gu, Seoul, Republic of
    Korea %
\item INAF -- Osservatorio Astronomico di Roma, via Frascati 33,
  I-00040 Monteporzio Catone, Italy %
\item Korea Astronomy and Space Science Institute (KASI), 776 Daedeokdae-ro, Yuseong-gu, Daejeon 305-348, Korea %
\item Department of Physics, Tokyo Institute of Technology, 2-12-1
  Ookayama, Meguro-ku, Tokyo 152-8551, Japan %
\item Department of Physics, Nihon University, 1-8-14 Kanda-Surugadai,
  Chiyoda-ku, Tokyo 101-8308, Japan %
\item W. W. Hansen Experimental Physics Laboratory, Kavli Institute
  for Particle Astrophysics and Cosmology, Department of Physics and
  SLAC National Accelerator Laboratory, Stanford University, Stanford,
  CA 94305, USA %
\item MAXI team, RIKEN, 2-1 Hirosawa, Wako, Saitama 351-0198 %
\item NASA Goddard Space Flight Center, Greenbelt, MD 20771, USA %
\item Oak Ridge Associated Universities %
\item Institute of Astronomy, National Central University, Chung-Li
    32054, Taiwan %
\end{affiliations}

\begin{abstract}
Massive black holes are believed to reside at the centres of most
galaxies.  They can become detectable by accretion of matter, either
continuously from a large gas reservoir or impulsively from the tidal
disruption of a passing star, and conversion of the gravitational
energy of the infalling matter to light.  Continuous accretion drives
Active Galactic Nuclei (AGN), which are known to be variable but 
have never been observed to turn on or off. %
Tidal disruption of stars by
dormant massive black holes has
been inferred indirectly but the onset of a tidal disruption event has
never been observed.  Here we
report the first discovery of the onset of a relativistic accretion-powered jet in the
new extragalactic transient, Swift J164449.3+573451.  
The behaviour of this new source differs from both theoretical models of
tidal disruption events and observations of the jet-dominated AGN
known as blazars.
These differences may stem from transient effects associated with the
onset of a powerful jet.  
Such an event in the massive black hole at the centre of our Milky Way
galaxy could strongly ionize the upper atmosphere of the Earth, if
beamed towards us.
\end{abstract}

Swift J164449.3+573451 (hereafter \transient) was discovered by the
\swift\cite{Gehrels04} Burst Alert Telescope\cite{Barthelmy05}  (BAT, 15-150 keV) 
in a 20-minute image of the sky that began at 12:57:45~UT on 28~March 2011\cite{Cummings11}.  
The \swift\ X-Ray Telescope\cite{Burrows05} (XRT) began observing the
field at 13:22:19.8~UT and found a bright, uncatalogued X-ray source.
The most accurate XRT position\cite{Goad07,Evans09} is RA (J2000): 
$16^{\rm h}~44^{\rm m}~49.92^{\rm s}$,
Dec (J2000): $+57^\circ~35'~00.6''$, with a 90\%
confidence error circle radius of 1.4 arcseconds.
A precise localization by the EVLA found a
variable source at the centre of a galaxy within the XRT error
circle\cite{Zauderer11}.  
Optical spectroscopy of
this source\cite{Levan11,Thoene11} measured a redshift of 0.354, corresponding to a
luminosity distance\footnote{assuming $H_0=71~{\rm km~s^{-1}},~\Omega_m=0.27,
 ~{\rm and}~\Omega_\Lambda=0.73$.} of 
$5.8 \times 10^{27}$ cm.

Subsequent analysis of BAT data taken before the on-board trigger
shows that the outburst was first detected on 
25 March 2011, with peak BAT count
rates of $\sim 0.02$ \batrate\ during the days just
before the on-board trigger (Supplementary Figure 1)\footnote{During the
brightest parts of the outburst, \transient\ had
power law photon indices ranging from 1.3 to 1.8 in the BAT bandpass;
the corresponding flux conversion factor is
$\sim 10^{-7}$ erg/count.  See Supplementary Table 1.}.  
Beginning on 28 March the BAT data show multiple flares peaking at up to 0.09 \batrate\
(about 20\% of the count rate from the Crab nebula).  
The source flux then dropped dramatically, with an average
count rate of  $0.0020 \pm 0.0005$ \batrate\  between 2
April 2011 and 18 April 2011.  

The flares seen by BAT are tracked in the 0.3-10 keV band by the XRT
(Figure~1, Supplementary Figures 1 and 2).   
Following nearly 3 days of intense flaring with
peak fluxes\footnote{All XRT fluxes are for the 0.3-10~keV band.} over $5 \times 10^{-9}$ \cgs, \transient\ decayed over
several days to a flux of about $8 \times 10^{-12}$ \cgs, then rose rapidly to
$\sim 8 \times 10^{-10}$ \cgs\ (Figure~2).  Over the next two
weeks the X-ray flux decayed gradually to $\sim 1.5 \times 10^{-10}$ \cgs, 
with occasional sharp dips to $\lesssim 3 \times 10^{-11}$ \cgs.
No statistically significant periodic or quasi-periodic signals were
found in the XRT data (see Supplementary Methods).

The X-ray spectral shape varies strongly with X-ray flux in the first
few days after the BAT trigger (Supplementary Figures 3 and 4).  A simple power law can fit the
spectra, but the fits are improved with one of several more complex models  %
(see Supplementary Methods for details).  Substantial
absorption in the host galaxy is required, with ${\rm N_H} \sim 2 \times 10^{22}$~cm$^{-2}$
(Supplementary Figure 5).

We obtained photometry in the uvw2, uvm2, uvw1, u, b, v, R, J, H, and
Ks bands, as well as broad-band ``white'' with the \swift\ UVOT, TNG,
BOAO, LOAO, and Maidanak Observatory telescopes (see Supplementary Methods for details).
The heavily obscured optical counterpart of the transient X-ray source
is not detected in optical or UV bands, where the host galaxy
dominates the light, but is detected strongly in the near-infrared.

Upper limits from the {\it Fermi} Large Area
Telescope\cite{Atwood09} (LAT) and from {\it VERITAS}\cite{Aliu11}
provide constraints
in the GeV and TeV band, respectively (see Supplementary Methods for details).

Archival X-ray and $\gamma$-ray data show no evidence of \transient\ before 25 March 2011.
X-ray flux upper limits from observations between 1990 and March 24
are orders of magnitude lower than the peak X-ray fluxes observed by
BAT and XRT, and are significantly lower than the lowest flux recorded
by XRT since \swift\ observations began on 28 March 2011 (see details
in Supplementary Methods).  The source has
brightened by at least 4 orders of magnitude compared with some
earlier deep observations of this field, and we consider that the
source was truly dormant for decades before this outburst began.

The abrupt appearance of a bright new persistent extragalactic X-ray source is
unprecedented. 
Gamma-ray bursts reach similar fluxes, but fade rapidly much more
rapidly than \transient.
Blazars (AGN with energetic jets pointed toward Earth) are known to be
highly variable, but are persistent and have never been observed to
turn off or to turn on, or display this level of variability.  
Supernovae in distant galaxies are rarely detected at X-ray
wavelengths.  %

\newcounter{savefn}
The isotropic X-ray luminosity\footnote{In the XRT band, which
is 0.4--13.5~keV in the \transient\ rest frame.}\setcounter{savefn}{\value{footnote}}
of \transient\ (corrected for absorption in the source)
ranges from $10^{45} - 4 \times 10^{48}$ erg~s$^{-1}$, with an
integrated X-ray power\footnotemark[\value{savefn}]
(over the 23 days following the BAT trigger) of 
$\sim 3 \times 10^{53}$~erg.
If \transient\ radiates isotropically, its total energy production is significantly larger than
the electromagetic energy released by a supernova explosion.
The persistence of the X-ray emission implies that this object is not powered
by an impulsive event such as an exploding star; instead, it requires
a continuous production of prodigious amounts of energy.
Only gravitational energy released by accretion of matter in the
vicinity of a black hole can produce such power levels over such a
long time period.
The observed minimum $3 \sigma$ X-ray variability time scale of \transient,
$\delta t_{\rm obs} \sim 100$~s, constrains the size of the black
hole under the assumption that the central engine dominates the
variability\footnote{Variability can also be produced in the jet, as
  is seen in blazars..}.
For a Schwarzschild black hole with mass $M_{\rm bh}$ and radius $r_s$, the minimum
variability time scale in its rest frame is
$\delta t_{\rm min} \sim  r_s/c \sim 10.0 (M_6)~{\rm s}$, where 
$M_6 = (M_{bh}/10^6 M_\odot)$. 
At $z=0.354$, this gives 
\begin{equation}
M_{\rm bh} \sim 7.4 \times 10^6 \left(\frac{\delta t_{\rm obs}}{100~{s}}\right)~M_\odot~.
\end{equation}
Much smaller masses are unlikely. 
The black hole mass can also be constrained through the $M_{\rm bh} -
L_{\rm bulge}$ relation,
which gives an upper limit of $2 \times 10^7~M_\odot$.
We conclude that the black hole mass is in the range $1 < M_6 < 20$ (see Supplementary Discussion).

The Eddington luminosity\footnote{The luminosity at which the
  gravitational force on infalling gas is balanced by outward radiation pressure.} of this black hole is
$L_{\rm Edd} = 1.3 \times 10^{44} ~{\rm erg~s^{-1}} M_6$~.
The observed peak luminosity is super-Eddington and requires a
strongly anisotropic radiation pattern with a relativistic jet pointed towards us.
A relativistic jet is also required by the radio
transient\cite{Zauderer11}, which we interpret as an external shock
in the gas surrounding \transient.

The spectral energy distribution (SED) produced from our data (Figure 3)
shows that the broad-band
energy spectrum is dominated by the X-ray 
band, which accounts for 
20\% (50\%) of the total bolometric energy
output in the low (high/flaring) X-ray state\footnotemark[\value{savefn}]. 
The SED places strong constraints on the emission
mechanism.  The IR-to-X-ray slope, nearly as hard as $\nu^{1/3}$, can
only be
accommodated within synchrotron theory in the absence of low-energy
electrons, which would severely overproduce the optical flux.  This
requires a particle-starved source, implying that the jet is
magnetically-dominated (see Supplementary Discussion for details, and
discussion of other models).  At high energies, the upper limits
from {\it Fermi} and {\it VERITAS} require that the self-Compton flux
from the X-ray peak is absorbed by the $\gamma$-$\gamma \rightarrow
e^{\pm}$ process.
The bulk Lorentz factor in the X-ray emitting
region must therefore be $\Gamma \lesssim 20$; we assume a value of $\sim 10$.

This luminous, accretion-powered, relativistic jet can be powered by
the tidal disruption of a star or by the onset of AGN activity from a
previously dormant massive black hole.
The \swift\ BAT, with a field of view of $\sim 4\pi/7$~sr, has
detected one such event in $\sim 7$ years at a flux that would have
been detectable to $z \sim 0.8$.  The all-sky rate of \transient-like
events is therefore $R_{\rm 4\pi}\sim 1~{\rm yr^{-1}}$, with a
90\% confidence interval\cite{Kraft91} of $0.08 - 3.9$~yr$^{-1}$.
The rate of tidal disruption events is thought to be 
$\sim 10^{-5}~{\rm yr^{-1}~galaxy^{-1}}$ on both
observational\cite{Gezari09} and theoretical\cite{Wang04} grounds.  
Given the galaxy number density
$n_{\rm gal} \sim 0.01 ~{\rm Mpc^{-3}}$, the total number of galaxies
in the co-moving volume within $z<0.8$ 
can be estimated as $N_{\rm gal}(z<0.8) \sim 10^9$. This gives a total
tidal disruption event rate of $R_{\rm TD,4\pi}(z<0.8) \sim
10^4~{\rm yr^{-1}}$. 
 If these act like AGN, 10\% of them contain relativistic
 (``radio-loud'') jets.  
The fraction of these jets that point towards us is
$$\max \left(\frac{1}{2\Gamma^2},
  \frac{\theta_j^2}{2} \right),$$ where $\Gamma$ is the bulk Lorentz
factor and $\theta_j$ is the opening angle of the jet.
A bulk Lorentz factor of $\Gamma \sim 10-20$ or a jet opening angle of
$\theta_j \sim 5^\circ$ would provide the observed event rate.
Our observations are consistent with theoretical predictions of the
formation of a a low-density,
magnetically-dominated jet during the super-Eddington phase of a tidal
disruption event around a $10^6 - 10^7 M_\odot$
black hole\cite{Strubbe09}.  
In the tidal disruption interpretation, we expect the source to begin
a slow decay once the material in the accretion disk is exhausted and
the luminosity begins to track the fall-back timescale ($t^{-5/3}$) of the
disrupted stellar material onto the black hole\cite{Lodato09}.  This decay may have
begun a few days after the BAT trigger (Figure~2).

A dormant massive black hole could also become visible through the onset of
long-lived AGN activity.  
The typical life time of an AGN is $t_{\rm AGN} \sim 3 \times 10^7$ 
yr\cite{Krolik99}.  In steady-state, the rate at which AGN turn on
is therefore $\sim t_{\rm AGN}^{-1}$ (but see Supplementary
Discussion). 
Within the $z<0.8$ volume, the event rate of
the onset of a new radio-loud AGN would be $R_{\rm AGN,tot} = 0.1 N_{\rm gal}(z<0.8) / t_{\rm AGN}
\sim 3~{\rm yr^{-1}}$. 
A moderate Lorentz factor of $\Gamma \sim 3$ or an opening angle of
$\theta_j \sim 13^\circ$ would have an
event rate of $\sim 0.1$~yr$^{-1}$, consistent with the
observations.  
The onset could be initiated by an accretion disk  in a rotating galactic bulge,
populated by gravitationally captured mass lost from stars in the
inner galaxy through supernovae or stellar winds.  The
inner radius of this disk will decrease slowly under the effect of $\alpha$
viscosity, with a final infall timescale of $t_d \sim 2 \times 10^3
(\alpha/0.1)^{-1} M_6$~s, resulting in a rapid onset of
electromagnetic radiation.
In this case one expects an extremely long-lived event
($\sim t_{\rm AGN}$).
Long-term monitoring of \transient\ will be required to distinguish
between these scenarios for the accretion.

We can consider \transient\ to be an analog of Sgr A*, the massive
black hole in the centre of our Milky Way galaxy.  Sgr A* has a mass
of $\sim 4 \times 10^6 M_\odot$\cite{Ghez08,Gillessen09}, and yet has
a quiescent soft X-ray
luminosity\cite{Baganoff01,Shcherbakov10} of only $4 \times 10^{32}$ erg
s$^{-1}$, which would be undetectable at z=0.354.  
If a jet similar to \transient\ turned on in Sgr A* and was pointed at
Earth, the peak X-ray flux
at the Earth could reach $\sim 200$~\cgs.  The X-ray flux would
correspond to a 
Y200-class solar flare, $> 40$ times
brighter than the strongest X-ray flare ever
detected from the Sun\cite{Thomson05}, and the fluence (integrated
flux) could exceed $10^4$~J~m$^{-2}$.
Such an event would produce strong ionization effects on the
Earth's atmosphere, and the total fluence is $>10$\% of that thought
necessary to produce mass extinctions 
in the case of a 10~s duration $\gamma$-ray burst\cite{Thomas05}.
Fortunately, the probability of a Sgr A* jet being pointed at Earth is low.

\bibliography{SwiftJ1644}

{\bf Supplementary Information} is linked to the online version of the paper
at www.nature.com/nature.

{\bf Acknowledgements}
DNB, JAK, and JMG acknowledge support by NASA contract NAS5-00136.
JPO, KLP, PAE, and APB acknowledge the support of the UK Space Agency. We
thank Dr.\ Andrew Read for help with the most recent XMM slew data.
SCa, SCo, PD'A, DF, GT, LAA, GC, AM, and BS  acknowledge financial support
from the agreement ASI-INAF I/004/11/0.
VM and PR acknowledge financial contribution from the agreement
ASI-INAF I/009/10/0.
PE acknowledges financial support from the Autonomous Region of
Sardinia under the program PO Sardegna FSE 2007--2013, L.R. 7/2007.
BZ acknowledges NSF (AST-0908362) and NASA (NNX10AD48G) for support.
ET was supported by an appointment to the NASA
Postdoctoral Program at the Goddard Space Flight Center,
administered by Oak Ridge Associated Universities
through a contract with NASA.
NK acknowledges support by the Ministry of Education, Culture, Sports,
Science and Technology (MEXT) through Grant-in-Aid for Science
Research 19047001.  
KH, NK, HN, MS, YUe, and RU thank the MAXI operation team.
MI, WK, and YJ acknowledge support from the CRI grant 2009-0063616,
funded by the Korean government (MEST).
KYH and YU acknowledge support from the NSC grants
99-2112-M-002-002-MY3 and 98-2112-M-008-003-MY3.
The $Fermi$ LAT Collaboration acknowledges support from a number of
agencies and institutes for both development and the operation of the
LAT as well as scientific data analysis. These include NASA and DOE in
the United States, CEA/Irfu and IN2P3/CNRS in France, ASI and INFN in
Italy, MEXT, KEK, and JAXA in Japan, and the K.~A.~Wallenberg
Foundation, the Swedish Research Council and the National Space Board
in Sweden. Additional support from INAF in Italy and CNES in France
for science analysis during the operations phase is also gratefully
acknowledged.
We acknowledge the use of public data from the \swift\ data archive.
This work also made use of data supplied by the UK Swift Science Data
Centre at the University of Leicester. 
The Fermi data are
available from http://fermi.gsfc.nasa.gov/ssc/.
We gratefully acknowledge the contribution of pre-publication upper
limits by the {\it VERITAS} Collaboration.

{\bf Author Contributions}
DNB, JAK, and ME composed the text, based on inputs from the other
co-authors.  Theoretical interpretation was provided by GG, BZ, ME,
and PM, with contributions by ADF and SC.  JAK, VM, KLP, JPO, PR, SCa, APB,
VD'E, PE, PAE, and GI
processed and analyzed the \swift\ XRT data.  TS, JRC, and HAK
processed and analyzed the \swift\ BAT data.  Optical/NIR data were
obtained with the TNG, BOAO, and LOAO telescopes and were provided, 
reduced, and analyzed by SCo, PD'A, DF, KYH, MI, YJ, Y-BJ,
W-KP, H-IS, GT, YUr, and LAA.  \swift\ UVOT data were processed and
analyzed by AAB, MMC, and STH with contributions by FEM.  {\it Fermi} LAT data analysis was
  performed by RC, NO, JSP, and ET.  KH, NK, HN, MS, YUe, and RU processed and
  analyzed the {\it MAXI} data. ADF provided liason with the
  {\it VERITAS} Collaboration.  JG and PG provided analysis of ROSAT
  archival data, and JO provided analysis of archival XMM data.  
All authors discussed the results and commented on the manuscript.

{\bf Author Information}
\swift\ data are available from the NASA HEASARC \\
(http://swift.gsfc.nasa.gov/docs/swift/archive/) or from mirror sites
in the UK \\
(http://www.swift.ac.uk/swift\_portal/archive.php) and 
Italy (http://swift.asdc.asi.it/).
Reprints and
permissions information is available at www.nature.com/reprints.
The authors have no competing financial interests.    
Correspondence
should be addressed to DNB (burrows@astro.psu.edu).

\begin{figure}
       \centering
        \parbox{3.5in}{
        \includegraphics[width=3.45in]{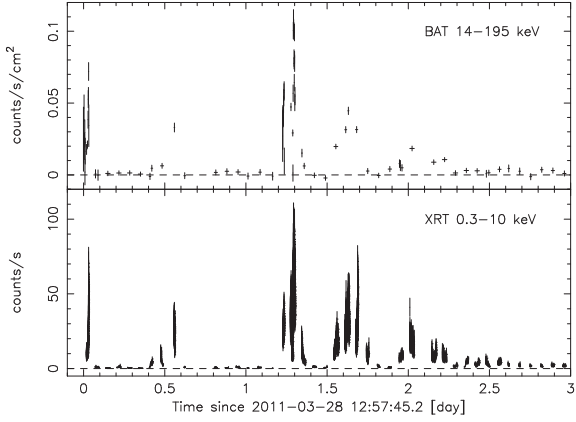}
        \centering\parbox{3.4in}{
\caption{\textbf{\swift\ BAT (top) and XRT (bottom) light curves of the early
  flares from \transient.}  The horizontal axis is time since the BAT
trigger on 28 March 2011.  The BAT and XRT count rates track each
  other, with episodes of flaring in the first
  few days after the BAT trigger on 28 March 2011.  Data gaps are caused by times
  when the source was not being observed.  Error bars are one standard
deviation.}}}
\end{figure}

\begin{figure}
       \centering
        \parbox{3.5in}{
        \includegraphics[width=2.5in,angle=270]{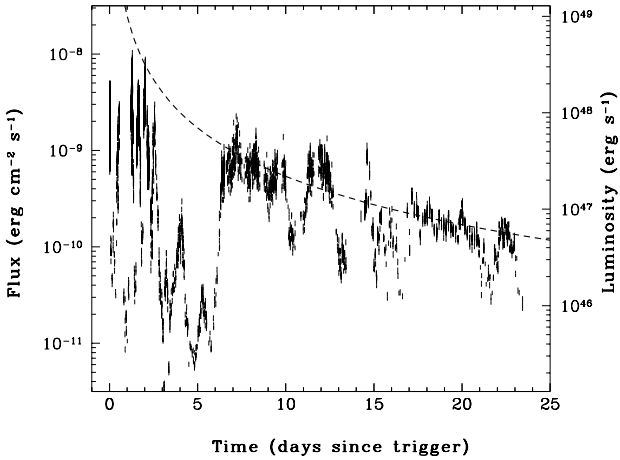}
        \centering\parbox{3.4in}{
\caption{\textbf{\swift\ XRT light curve of \transient\ for the first 3 weeks
  of observations.}   The X-ray events were summed into time bins containing 200
  counts per bin and count rates were calculated for each time bin.
Because of the strong spectral evolution of this source and its large
intrinsic absorption, light curves in physical units cannot be 
produced with constant count rate-to-flux conversion factors.
We use time-dependent
  conversion factors to convert count rates to fluxes.
  Spectral fits were performed on time slices of data that typically
  contained about 3000 counts (see Supplementary Methods).  
Count rate-to-flux conversion factors
  were calculated for each spectrum, using absorption-corrected fluxes
  in the rest frame 0.4-13.5 keV band.  
These time-dependent flux conversion factors were then
  interpolated to obtain fluxes for each time bin.  
The right-hand axis gives the conversion to luminosity
  of the source, assuming isotropic radiation.  The dashed line shows
  a $t^{-5/3}$ curve (where $t$ is time since the BAT trigger).  Error bars are one standard
deviation.}}}
\label{fig:xrt_lc}
\end{figure}

\begin{figure}
       \centering
        \parbox{3.5in}{
        \includegraphics[width=3.4in]{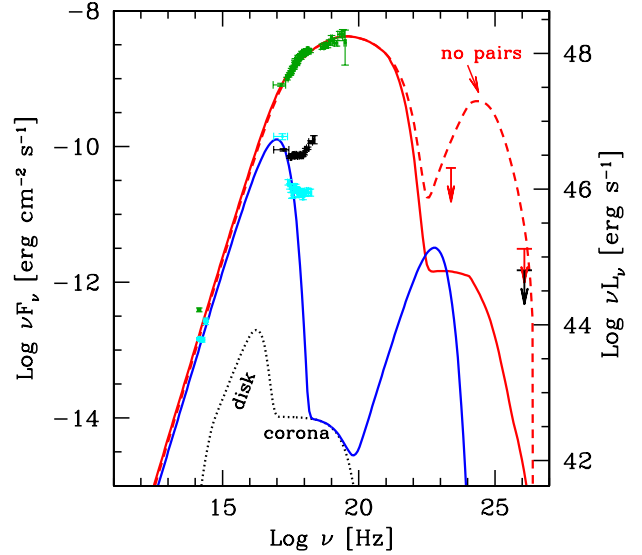}
        \centering\parbox{3.4in}{
\caption{\textbf{The Spectral Energy Distribution (SED) of \transient.}  The green
data points show the transient brightness a few days after the BAT
trigger; cyan data points show the measured SED at 4.5 days; black
data points are XRT data between 6.5 and 9.5 days after the BAT trigger.  
Near-infrared (NIR) were obtained with the TNG and BOAO
telescopes in the Ks
band 48 hours after the trigger, and in the J, H,
and Ks band fluxes 4.5 days after trigger. 
The NIR flux has been dereddened with $A_V=4.5$.
Details of the observations and data reduction are presented in
Supplementary Methods.
In the X-ray band we show the spectrum at the peak of the bright flares
(31 hours after BAT trigger; XRT and BAT) and in the low flux state ($\sim
4.5$ days after, XRT only), 
together with the spectrum in the intermediate persistent 
flux state that began a week after the BAT trigger
(time integrated between day 6.5 and 9.5 from trigger, XRT only).
The X-ray data have been corrected for absorption with a constant $N_H=2
\times 10^{22}$~cm$^{-2}$. %
Upper limits from the {\it Fermi} LAT at $2 \times 10^{23}$~Hz and
from {\it VERITAS}\cite{Aliu11} at $10^{26}$~Hz are also shown.
The red curve shows the model discussed in the text, which
is a blazar jet model\cite{Ghisellini09} fit to our SED.  The
dominant emission mechanism is synchrotron radiation peaking in the
X-ray band.  On the low frequency side, the hard slope between the
NIR and X-ray bands requires suppression of low-energy
electrons, which would otherwise overproduce the NIR flux.  
On the
high frequency side, the LAT
and {\it VERITAS} upper limits require that the self-Compton component is
suppressed by $\gamma$-$\gamma$ pair production.  
The model includes a disk/corona component from
the accretion disk (black dotted curve), but the flux is dominated at 
all frequencies by the synchrotron component from the jet.  
The blue curve shows the corresponding model in the low X-ray flux state. 
The kink in the X-ray spectrum suggests that a possible additional component may be
required; it would have to be very narrow, and its origin is unclear.
The model parameters for this fit are given in the Supplementary
Discussion, and suggest a magnetically-dominated jet, which is
unusual for AGN.   This might be explained by the unusually high
accretion rate and small black hole mass (compared to typical AGN) implied by our data.  
In the SI we discuss two alternative models: one with a
magnetically-dominated jet but a more AGN-like
parameter set, and one at the opposite extreme of a particle-dominated
jet.
Both of these alternative models seem contrived, and we favor the
model presented here.  All error bars are one standard
deviation.
}}}
\end{figure}

\end{document}


\title{Discovery of the Onset of Rapid Accretion by a Dormant
Massive Black Hole: Supplementary Information}

\maketitle

\section*{Supplementary Methods}

\subsection*{\swift\ Discovery and broad-band observations}

\paragraph{BAT data}

The \swift\cite{Gehrels04} Burst Alert Telescope\cite{Barthelmy05}
(BAT, 15-150 keV) triggered on a new uncatalogued source at $T_0 =$
12:57:45 UT 28 March 2011\cite{Cummings11}.  
The trigger was assumed to be a $\gamma$-ray burst (GRB) and was named
GRB\ 110328A, following standard nomenclature.
However, this trigger was followed by
three additional BAT triggers at increasing count rates over the next
two days\cite{Sakamoto11} (see Supplementary Table~\ref{tab:bat_triggers}), 
making it clear that this object was not,
in fact, a GRB.  
(Following the fourth trigger, the on-board BAT source catalog was
adjusted to prevent further triggers.)
The X-ray counterpart was renamed Swift J164449.3+573451, which
is now the preferred name of the source\cite{Barthelmy11}.
We will refer to Swift J164449.3+573451 as \transient\ for the
remainder of this paper.

\begin{table}
\centering
\caption{\textbf{\swift\ BAT trigger details for \transient.}}
\begin{tabular}{lllll}
\\
\hline
Trigger Number & Date & Time & Trigger Duration & Intensity \\
                    & & UT & (s) & (counts s$^{-1}$) \\
\hline
450158  &  28 Mar & 12:57:45  & 1208    &   6.1 \\
450161  &  28 Mar & 13:40:41   &  64     &  19.4 \\
\multicolumn{5}{l}{\qquad Threshold set to 0 so BAT would trigger on
  it again} \\
450257   & 29 Mar & 18:26:25   & 320   &   15.6 \\
450258   & 29 Mar & 19:57:45   &   64   &   38.2 \\
\multicolumn{5}{l}{\qquad Triggers from this source disabled} \\
\hline
\end{tabular}
\label{tab:bat_triggers}
\end{table}

Post-facto examination of pre-trigger data
indicate that the outburst was first detected by BAT on 25 March
2011 with a mean count rate (integrated over 24 hours) of ($0.0059 \pm
0.0016$) \batrate\ (too low to generate an on-board
trigger).  
We note that the source was outside the BAT FOV for 2.86 hours before
the first trigger. 
Between 25 March and 31 March the BAT data show multiple flares
peaking at up to 0.09 \batrate\ (about 22~mCrabs).  
During the brightest parts of the
outburst, the source had power law photon indices ranging from 1.3 to
1.8.  The average 15-150 keV flux in the time interval 12:57:45 UT on
28 March 2011 to 05:30 UT on 30 March 2011 was about $(1.0 \pm 0.2)
\times 10^{-9}$ \cgs.  The source flux then dropped
dramatically, with an average count rate of 
$0.002 \pm 0.0005$ \batrate\ between 2 April 2011 and 12 April 2011 (see
Supplementary Figure~\ref{fig:bat_xrt_lc}).

For most of the BAT data the spectral slope cannot be determined, making it
difficult to determine an accurate flux based on the count rates. However
if we assume a spectral slope, we can obtain estimates of the 15--150
keV flux. Supplementary Table~\ref{tab:bat_flux_conversion} provides 
flux conversion factors for average BAT count rates at 3 epochs, for 2
observed spectral slopes.

\begin{table}
\centering
\caption{\textbf{Flux conversion for observed BAT count rates for
    different spectral slopes}}
\begin{tabular}{llll}
\\
\hline
Time Period & BAT count rate & Photon Index & Estimated Flux \\
                    & (cts s$^{-1}$ cm$^{-2}$)&(assumed) & (\cgs, 15-150 keV)\\
\hline
Pre-Trigger & 0.0059 & 1.8 & $5.1\times10^{-10}$ \\
                   & 0.0059 & 1.3  & $6.1\times10^{-10}$ \\
Peak rate     & 0.0900 & 1.8 & $7.8\times10^{-9}$ \\
                   & 0.0900 & 1.3 & $9.3\times10^{-9}$ \\
Late time     & 0.0020 & 1.8 & $1.7\times10^{-10}$ \\
                   & 0.0020 & 1.3 & $2.1\times10^{-10}$ \\
\hline
\end{tabular}
\label{tab:bat_flux_conversion}
\end{table}

\begin{figure}[t]
       \centering
        \parbox{6.5in}{
        \includegraphics[width=6in]{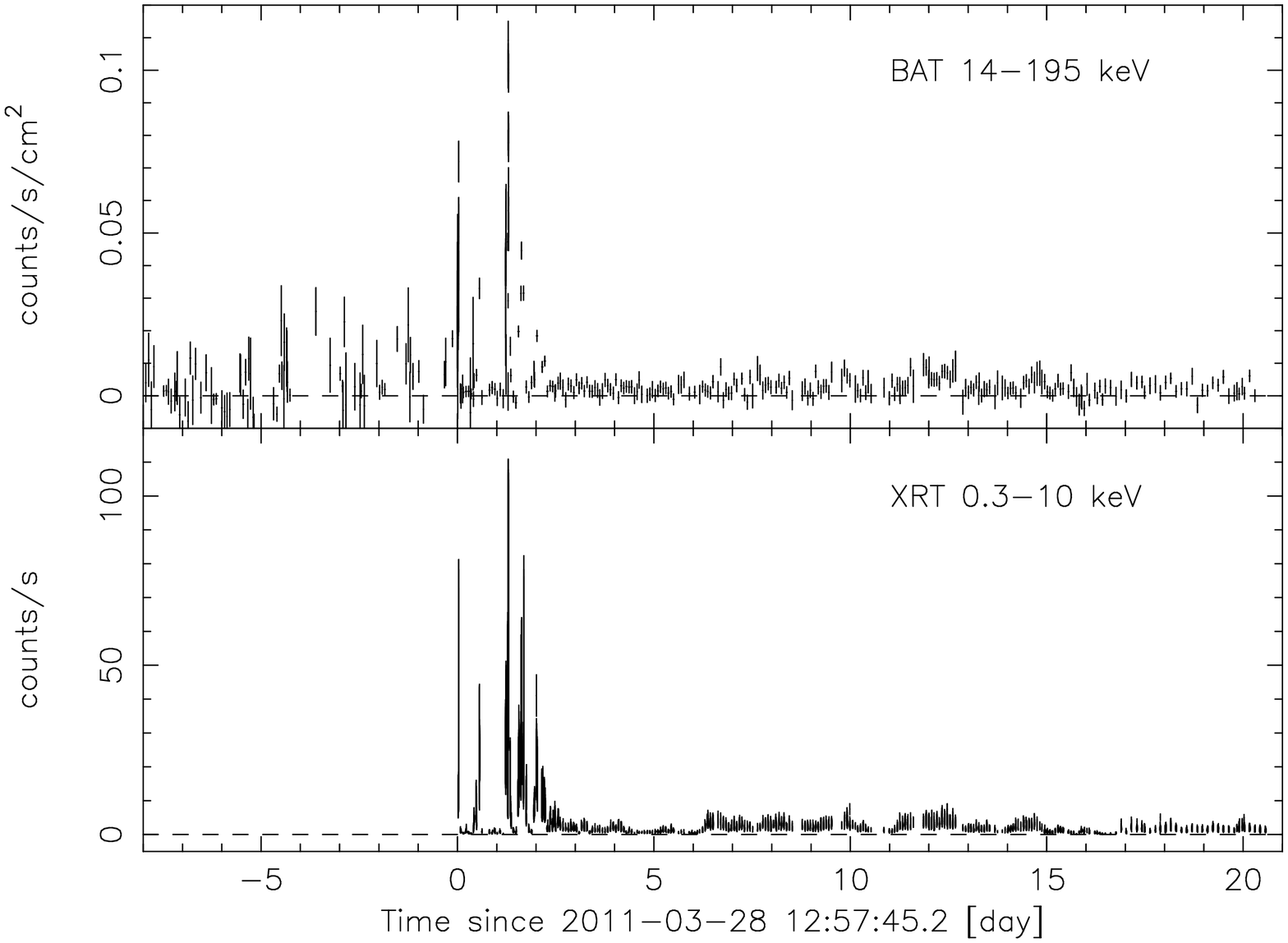}
        \centering\parbox{6in}{
          \caption{\label{fig:bat_xrt_lc}\textbf {BAT (top) and XRT (bottom) light curves of
              \transient.}  Times are measured from the BAT trigger on
            28 March 2011.  Although the higher count rate in the
            XRT allows better time resolution, it is clear that the
            count rates in both instruments track each other, with
            numerous flares in the first few days after the BAT
            trigger. Note that \transient\ was detected by BAT $\mathsf{\sim}$4$-$5
            days before BAT triggered and XRT observations
            began. Data gaps are caused by periods when the source was
            not being observed.}}}
\end{figure}

\paragraph{XRT data} 

Observations of \transient\ with the \swift\ X-ray Telescope
\cite{Burrows05} (XRT, 0.3-10 keV) began at 13:20:52 UT on 28 March
2011. XRT observed \transient\ daily for between 11\,ks and 28\,ks per
day. We report here on the first 23 days after the trigger,
during which the mean time spent on-target was 15.2 ks per day. 
As \swift\ is in a low-Earth
orbit, observations are broken into small snapshots, typically 20--30\,min
long per $\sim96$\,min \swift\ orbit.

For the first 14.5 days the XRT was in ``Auto State'', in which it
autonomously selects the appropriate observing mode based upon the
brightness of the observed source. This meant that as the source
varied in brightness, the XRT collected data in a combination of
Windowed Timing (WT) and Photon Counting (PC) modes. PC mode is the
standard XRT imaging mode, with a timing resolution of 2.5 s; if a
source is brighter than $\sim 0.5$ XRT count s$^{-1}$ in this mode, pile-up correction
must be used \cite{Vaughan06}. WT mode is a fast timing mode (1.8 ms time
resolution) that collects 1-dimensional image data. The fast readout
of WT mode means that pile-up is avoided for sources with a count rate
below $\sim 100$ counts\,s$^{-1}$. 

The switch points for Auto State are tuned specifically for GRB
observations, i.e. a fading X-ray light curve.  For
sources of moderate brightness (1-5 XRT counts s$^{-1}$), data will often be
collected in PC mode. For this reason we changed our observing mode to
WT for data collected from 12 April 2011 onwards, to avoid issues
relating to pile-up.

X-ray spectra and light curves were produced utilizing the methods
described by Evans et al. \cite{Evans09} 
All PC mode data were corrected for pile-up by
removing events from the core of the PSF, and utilizing the wings of
the PSF for both spectral fitting and light curve generation
\cite{Vaughan06}. 
XRT spectra were corrected for the effects of charge traps that have
developed due to radiation damage to the CCD over the \swift\ mission
lifetime. 

The XRT count rate light curve for \transient\ is shown in
Supplementary Figure~\ref{fig:xrt_lc}.  Total exposure time for the
first 23.4 days is $3.56\times10^5$ s, with a live time fraction of
17.6\%.

\begin{figure}[t]
       \centering
        \parbox{6.5in}{
        \includegraphics[width=6in]{xlcv_cr.eps}
        \centering\parbox{6in}{
          \caption{\label{fig:xrt_lc}\textbf{\swift\ XRT  count  rate  light  curve  of
              \transient, through 14  April 2011.}  Following nearly 3
            days of intense  flaring with peak count rates  of over 80
            cts s$\mathsf{^{-1}}$  , the source  decayed for several
            days to a count rate of about 0.7 cts s$\mathsf{^{-1}}$,
            then rose to $\mathsf{\sim}$2 cts s$\mathsf{^{-1}}$, where it has
            remained (with excursions) for over 9 days. From
            $\mathsf{\sim}$13 days after the BAT trigger, the
            light curve can be described by a mean level of
            $\mathsf{\sim}$1 cts s$\mathsf{^{-1}}$, with episodes of
            dipping that last 1$-$2 days where the count rate
            drops to as low as 0.2 cts s$\mathsf{^{-1}}$.
          }
}}
\end{figure}

\begin{figure}[t]
       \centering
        \parbox{6.5in}{
        \includegraphics[width=6in]{rat_softTot_thruT0+2164ks.eps}
        \centering\parbox{6in}{
          \caption{\label{fig:xrt_ratio}\textbf{Ratio of XRT count
              rates in the 0.3--1.3 keV and 0.3--10 keV bands, plotted
              against the overall XRT count rate.}  The band ratio is
            strongly correlated with teh overall count rate, 
            demonstrating that the spectrum gets progressively
            harder as the flux increases.  There is a discontinuity
            around 4 ct/s; data at higher count rates were collected
            during flares in the first few days of monitoring, whilst
            data at lower count rates were collected after the XRT
            observing mode was changed from Auto to WT 14.5 days after
            the BAT trigger.  The trend for harder spectra at higher
            count rates is seen on both sides of this discontinuity.
            This figure includes only WT-mode data taken within 23 days
            of the first BAT trigger.  Each data point averages all WT
            data collected within a single \swift\ orbit.  Only data points with uncertainties smaller than 25\% are shown.}}}
\end{figure}

We produced light curves in two energy bands (soft and hard) in order to
examine spectral variations with time and flux.
We observe a strong anti-correlation in WT mode data between the
spectral hardness and flux, with softer spectra as the source intensity
decreases (see Supplementary Figure~\ref{fig:xrt_ratio}).  There is a
similar correlation in PC mode data, but the scatter is larger and the
correlation is weaker.

\paragraph{UVOT data}

The \swift\ UV-Optical Telescope\cite{Roming05} (UVOT) observed
\transient\ in all optical and ultraviolet filters immediately
following each of the four \swift\ BAT triggers, switching to mainly the
ultraviolet filters for deep imaging after 1 April 2011.  Observing periods were the same
as for XRT.  No persistent source has been detected in any filter.
Photometry has been performed on individual and summed images using
the UVOT photometric system\cite{Poole08}.  No correction has been
applied for Galactic extinction due to the reddening of
$E_{B-V}=0.024$ in the direction of the transient\cite{Schlegel98}.
Supplementary Table~\ref{tab:uvot_ul} provides the upper limits for
two epochs: early X-ray variability (first two days after the
\swift\ BAT trigger), and later times.

\begin{table}
\centering
\caption{\label{tab:uvot_ul}\textbf{\swift\ UVOT UV upper
    limits.}}
\begin{tabular}{lllll}
\\
\hline
      & $T_0+0-2$ days & & $T_0 + 2-15$ days & \\
Filter & Exposure (s) & UL ($3\sigma$) & Exposure (s) & UL ($3\sigma$) \\
\hline
white & 18359 &  $>$23.6 & 11221 & $>$23.5 \\
v     &  2541 &  $>$21.1 & {\it no observations} & \\
b     &  1493 &  $>$21.5 & {\it no observations} & \\
u     & 31593 &  $>$23.1 & 37032 & $>$23.2 \\
w1    &  3422 &  $>$21.8 & 82742 & $>$23.5 \\
m2    &  3048 &  $>$21.6 & 38324 & $>$23.1 \\
w2    &  2320 &  $>$21.8 & 24763 & $>$23.1 \\
\hline
\end{tabular}

\end{table}

\subsection*{XRT Timing analysis}

After applying the barycentric correction to the XRT event lists we
searched for coherent or quasi-periodic signals over different energy
ranges and time intervals by using Fourier techniques. Due to the presence
of strong non-Poissonian noise, mainly introduced by the rapid
variability of the source, particular care must be taken in evaluating
the statistical significance of any candidate signal. Following the
prescriptions of Israel \& Stella \cite{Israel96}, no significant
(periodic or quasi-periodic) signal was found. Conservative 3$\sigma$ upper
limits on the pulsed fraction, defined as semi-amplitude of the sinusoid
divided by the mean source count rate, were computed. For the data span in
the 31 March 2011 to 19 April 2011 interval (i.e. without the initial intense spikes in
order to mitigate the low frequency noise in the power spectrum) and the
energy range $0.3-10$ keV (524288 trials between 1$\times$10$^{-6}$ and 0.2
Hz), the 3$\sigma$ upper limits are in the 1\%-3\% range for periods
shorter than 500s, and in the 5\%-90\% range for periods between $\sim$500s and 1000s.

\subsection*{X-ray spectral analysis} 

The XRT data were processed with standard procedures ({\sc
  xrtpipeline} v0.12.4), filtering and screening criteria by using
{\sc ftools} in the {\sc Heasoft} package (v.6.9). We used the latest (pre-release)
version of the XRT software task {\em xrtcalcpi}, which calculates
Pulse Invariant (PI) event energies taking into account position
dependent corrections for energy losses incurred by charge traps that
have developed on the CCD. These software and calibration files are
only available to the XRT team at the time of this
analysis, but will be released to the public by the time this paper is
published.  %
They are the result of several years of calibration efforts by the XRT
team to correct for the effects of radiation damage to the XRT
detector after $>6$ years in orbit. They provide significant
improvements in the spectral resolution and flux calibration compared
with earlier versions.  We used the latest pre-release spectral
redistribution matrices (swxwt0to2s6\_20010101v013.arf and
swxwt0to2s6\_20070901v013.rmf for WT data and
swxpc0to12s6\_20010101v012.arf and swxpc0to12s6\_20070901v012.rmf for
PC data).

\paragraph{Intensity-selected spectra}
We produced XRT spectra in seven intensity ranges for data collected 
from March 28 until April 7 2011. Events were accumulated with count rates in the
following intervals $<0.5$, $0.5-1$, $1-2.5$, $2.5-10$ counts s$^{-1}$ for the
PC data, and $<11$, $11-35$, $>35$ counts s$^{-1}$ for WT data.  
We corrected the PC data for pile-up 
by determining the size of an exclusion region at the core of the PSF
necessary to get agreement between the wings of the
observed PSF and the nominal PSF \cite{Vaughan06}, and excluding from the
analysis all the events that fell within that region. 

The spectra were fit with several models. In all cases the absorption
consisted of two components, one fixed to the Galactic value
($N_{\rm H}=1.7\times10^{20}$ cm$^{-2}$), and one free at the redshift $z=0.35$.
The following models provided good fits to the data:
{\it i)} a simple absorbed power-law model (tbabs*zwabs(powerlaw)); 
{\it ii)} a log-parabola model (tbabs*zwabs(power2)); 
{\it iii)} a broken power-law model (tbabs*zwabs(bknpower)).
{\it iv)} an absorbed power-law model plus a diskblackbody (tbabs*zwabs(powerlaw+diskbb)).
The ``power2'' log-parabola model is defined as:
\begin{equation}
A(E) = E^{(-\alpha+\beta*\log E)}
\label{eq:logparabola}
\end{equation}
The spectra are comparably well-fit well by either a log-parabola, a
broken power-law or power-law plus multi-temperature thermal model.
Again, we observe a trend for softer spectra as the source intensity
decreases (see Supplementary Figure~\ref{fig:xrt_spectral_evolution_7}).

\begin{figure}[t]
      \centering
       \parbox{6.5in}{
       \includegraphics[height=6in,angle=270]{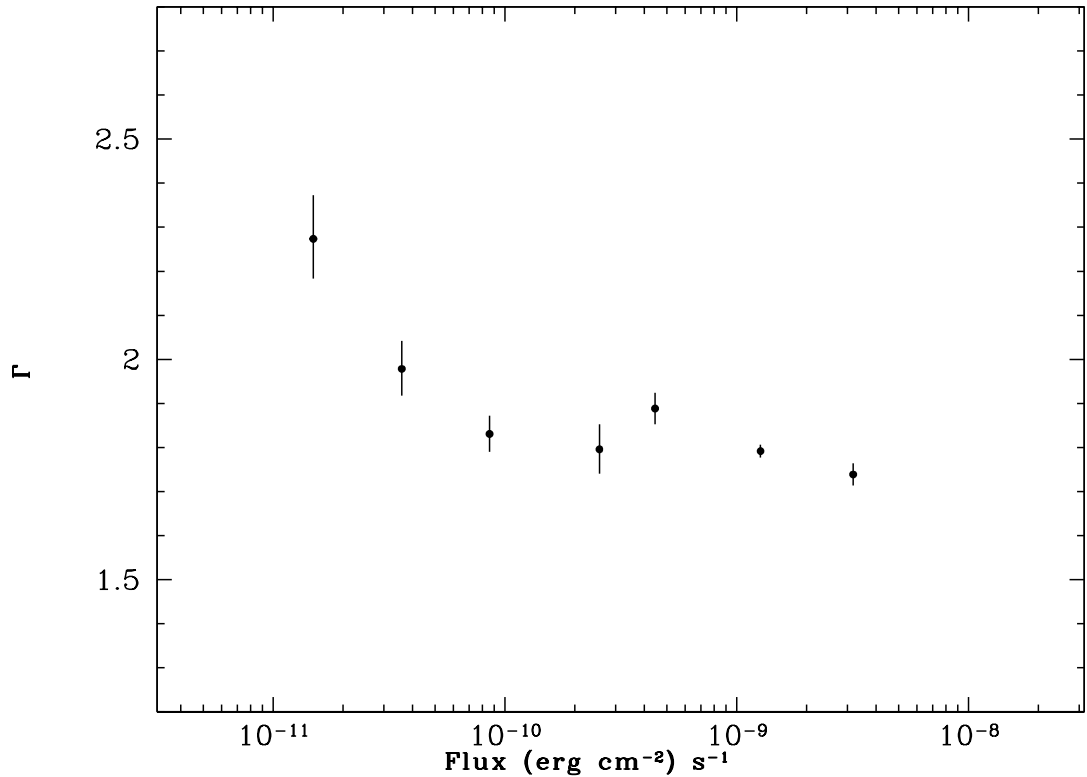}
       \centering\parbox{6in}{
\caption{\label{fig:xrt_spectral_evolution_7}\textbf{\swift\ XRT
    intensity-resolved spectral fitting results
utilizing an absorbed power-law model.}  Flux is for the 0.3--10 keV
band.
The power law slope is
strongly correlated with the count rate, confirming the
band ratio results shown in Supplementary Figure~\ref{fig:xrt_ratio}. 
}}}

\end{figure}

\paragraph{Time-selected spectra}
Spectra were also accumulated from all available XRT data with the aim of
obtaining at least 3000 counts per spectrum, unless relevant intensity
variations were observed, in which cases more spectra were extracted, to
represent the different intensity states.  Hence some spectra were also
extracted with fewer counts than 3000.  
We obtained a total of 32 PC spectra and 98 WT spectra, and we fit them with 
the first three models described above. 
The simple power-law model is an adequate description of the data, 
as is the log-parabola (F-test probabilities $\sim 10^{-3}$).
We find that the column density is variable in time
with no evidence of increased column when the dips occur. 
The results of the temporally resolved log-parabola model fits are
shown in Supplementary Figure~\ref{fig:xrt_spectral_evolution}; in particular, 
the harder when brighter trend is also observed (the average photon index 
during the first three days and on the 15th day is $1.78\pm0.02$ and $1.91\pm0.08$,
respectively). 

\paragraph{Flux-calibrated XRT light curves}
The results of the spectral analysis with the log-parabola fits were
used to calculate energy correction factors (ECFs) for conversion
from the XRT count rate light curve shown in Supplementary
 Figure~\ref{fig:xrt_lc} to the flux-calibrated light curve
presented in Figure~2 of the main Letter.  The absorption-corrected
fluxes (corresponding to the observed fluxes in 
Supplementary Figure~\ref{fig:xrt_spectral_evolution}) were used to
calculate ECFs corresponding to each time-resolved spectrum.  These
ECFs were then
interpolated to the time of each bin in the count rate light curve of 
Supplementary Figure~\ref{fig:xrt_lc}.  The interpolated ECFs were
multiplied by the counts in each light curve bin to obtain the
absorption-corrected flux light curve.

The flux light curve was used to estimate the total X-ray energy
produced by \transient.  The total measured unabsorbed fluence is $1.24\times10^{-4}$ erg
cm$^{-2}$.  Correcting for the live-time fraction, this gives a total
unabsorbed fluence for \transient\ of $7.1 \times 10^{-4}$ erg
cm$^{-2}$ in the observed 0.3--10.0 keV band. 
Each flux point in the light curve was also converted to a luminosity
at the source, and these were summed to obtain an estimate of the
total X-ray energy (corrected for live-time fraction) of
$3 \times 10^{53}$ erg in the 0.4--13.5~keV rest frame energy band. 
Mean, median, peak and minimum flux and luminosity values for
\transient\ are given in Supplementary Table~\ref{tab:flux_levels}.

\begin{table}
\centering
\caption{\label{tab:flux_levels}\textbf{X-ray Rest Frame Flux and Luminosity of
    \transient.}}
\begin{tabular}{lcc}
\\
\hline
& Observer frame & Rest frame \\
Flux Level & Unabsorbed Flux (0.3--10.0 keV) & Luminosity (0.4--13.5 keV) \\
\hline
Max flux&	$1.03\times10^{-8}$	\cgs       &
$4.30\times10^{48}$	$\mathrm{erg\ s^{-1}}$\\
Mean flux&	$1.35\times10^{-9}$	\cgs     & 	$5.63\times10^{47}$	$\mathrm{erg\ s^{-1}}$\\
Median flux&	$7.07\times10^{-10}$	\cgs         & 	$2.96\times10^{47}$	$\mathrm{erg\ s^{-1}}$\\
Min flux&	$3.35\times10^{-12}$	\cgs          &   	$1.40\times10^{45}$	$\mathrm{erg\ s^{-1}}$\\
\hline
\end{tabular}
\end{table}

\paragraph{XRT spectra for SEDs}
We also extracted spectra for use in our SED during three epochs representing three intensity states: 
{\it i)} strictly simultaneous with the event-mode BAT spectrum; 
{\it ii)} a very low state (4.5--5.1 days after the trigger);
{\it iii)} an intermediate state (6.5--9.4 days after the trigger). 
The XRT data for these states are plotted in Supplementary
Figures~\ref{F:sed-one}--\ref{F:sed-three}.

\begin{figure}[t]
       \centering
        \parbox{6.5in}{
        \includegraphics[height=6in,angle=270]{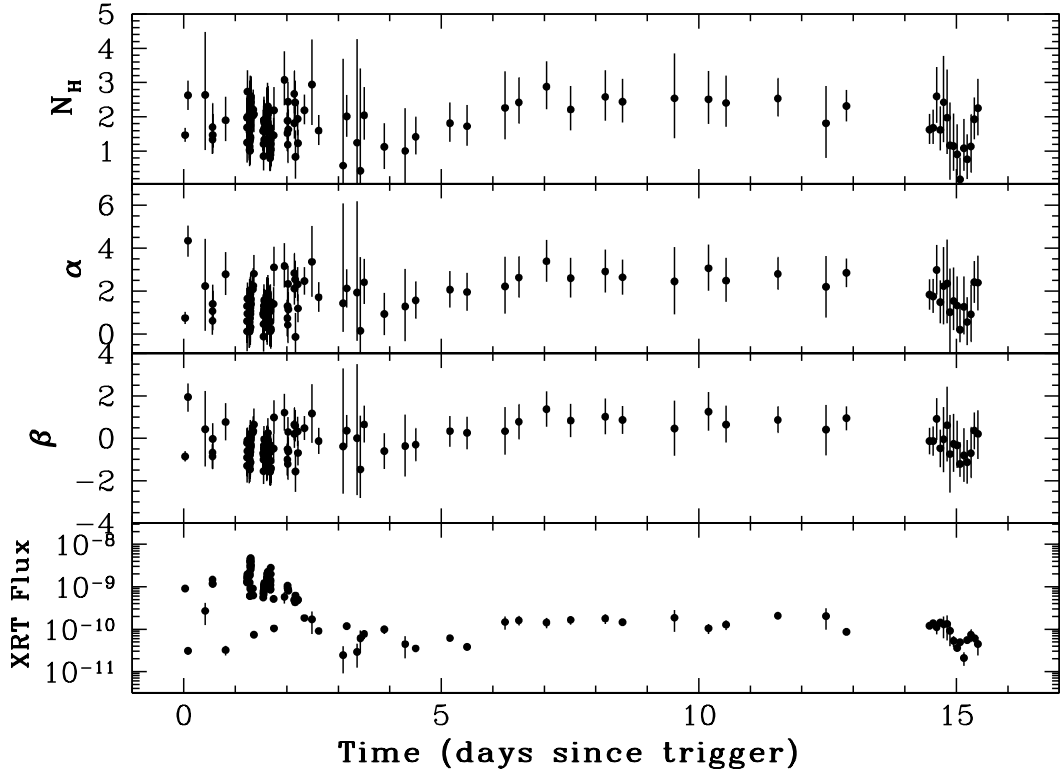}
        \centering\parbox{6in}{
\caption{\label{fig:xrt_spectral_evolution}\textbf{\swift\ XRT time resolved spectral fitting utilizing an
    absorbed log-parabola model.} Parameters $\alpha$ and $\beta$ for
  the log-parabola model are as defined in Supplementary
  Equation~\ref{eq:logparabola}. $N_{\rm H}$ is the redshift corrected
  intrinsic absorption in units of $\mathsf{10^{22}}$
  cm$\mathsf{^{-2}}$, XRT flux is the observed flux in
  units of \capcgs (0.3$-$10 keV).}}}

\end{figure}

\paragraph{Simultaneous BAT+XRT spectral fits}
The results of fitting the BAT+XRT spectral data simultaneously for
one of the early flares can be seen in Supplementary
Figure~\ref{fig:bat_xrt_spec}. We find that for this time period the
BAT and XRT spectral fit are consistent, fitting well a broken
power-law model.%
The BAT $15-150$ keV spectrum is a continuation of the XRT
spectrum, and there is no need for additional spectral breaks or high
energy components to fit the BAT spectrum.
Unfortunately it is not possible to fit perform simultaneous fits at
all epochs, due to source faintness and the limitations of the Survey
data collection mode of the BAT during regular observations. The
spectral parameters appear to be strongly dependent on the X-ray flux
(see Supplementary Figures~\ref{fig:xrt_ratio}--\ref{fig:xrt_spectral_evolution_7}). 
Therefore in order to
test whether the BAT spectra were typically consistent with the XRT fits,
we fit a broken power-law model to the XRT spectrum at various flux
levels, and then extrapolated this model to estimate the expected
count rate seen in the BAT energy band. 
We found that the BAT count rates are highly consistent with the
extrapolated XRT spectral fits, suggesting that the consistency seen
in the BAT+XRT spectral fit show in Supplementary
Figure~\ref{fig:bat_xrt_spec} is true also for later time data.

\begin{figure}[t]
       \centering
        \parbox{6.5in}{
        \includegraphics[width=6in]{grb110328A_BAT_XRT.eps}
        \centering\parbox{6in}{
\caption{\label{fig:bat_xrt_spec}\textbf{\swift\ combined BAT/XRT ``unfolded'' spectra for \transient.}
  The black points are XRT data and the red points are BAT data.  The
  data are from the time interval 19:57:52.0 -- 20:04:32.0 UT on 29
  March 2011, during one of the bright flares.  This figure shows the
  spectrum incident on the instruments. The solid line is the fitted
  broken power-law model.  The spectra are equally well-fit by a
  combination of a multi-temperature disk and a power law.}}}

\end{figure}

\subsection*{Observations by other X-ray/$\gamma$-ray observatories}

\paragraph{\textit{MAXI}}

Following the BAT on-board trigger, the \textit{MAXI}\cite{Matsuoka09}
team reported that the Gas Slit Camera\cite{Mihara11} 
also detected \transient\cite{Kimura11}. 
The source rises to a peak
about a day after the first BAT trigger.  {\it MAXI} continued to
detect the source for the initial bright flaring part of the
outburst. The \textit{MAXI} one-day-averaged light curve for the period before and
after the BAT trigger is shown in Supplementary Figure~\ref{fig:maxi_outburst_lc}.

\begin{figure}[t]
       \centering
        \parbox{6.5in}{
        \includegraphics[width=6in]{maxi_lc.eps}
        \centering\parbox{6in}{
\caption{\label{fig:maxi_outburst_lc}\textbf{\textit{MAXI} one-day-averaged light curve covering the $\sim$20 days before and after the
  detection of \transient\ by BAT.}
}}}

\end{figure}

\paragraph*{\fermi}

In the GeV energy range, no significant $\gamma$-ray emission has been
seen by the \fermi\ Large Area Telescope (LAT) from the direction of
\transient. During the period of main activity (28 March to 4 April
2011) \fermi\ LAT observations constrain the average $\gamma$-ray flux to
$< 2.7\times10^{-11}$ \cgs (100 MeV-10 GeV, 95\% confidence
upper limit).  During the following week, the equivalent upper limit is 
$2.9 \times 10^{-11}$ \cgs.  Daily upper limits during the X-ray
outburst vary, but are typically a few times %
$10^{-10}$ \cgs.  For the LAT analysis the event class used was
``DIFFUSE'' and the version of the instrument response functions used
was Pass 6, v3 (P6\_V3\_DIFFUSE).

\subsection*{UV/Optical/NIR observations}

\transient\ has been observed by a large number of ground-based optical and
radio telescopes.  In general, the heavily extincted optical counterpart
is not clearly detected in optical or UV bands, where the host galaxy
dominates the light, but is detected strongly in the near-infrared.
Spectroscopic observations identified emission lines at a redshift of
0.354\cite{Levan11,Thoene11}.  Here we report details of optical and
NIR observations obtained by our team.

\paragraph{R-band Observations}

$R$-band imaging data were taken using a CCD camera on the
Mt. Lemmon Optical Astronomy Observatory (LOAO) 1-m telescope\cite{Han05,Lee10} in
Arizona, US, and SNUCAM on the Maidanak Observatory
1.5-m telescope in Uzbekistan.. The data were taken during the nights
of 29 March 2011 to 8 April 2011 at LOAO, and 12 April at Maidanak. 
A dithered sequence of 300\,s -- 600\,s exposures
were taken, resulting in 20\,min -- 1\,hour total integration for each night,
which gave $3\sigma$ detection limits of $R=22.4 - 23.9$ mag over a $3''$
diameter aperture. The photometric calibration %
is based on field calibration performed using four Landolt
standard star fields (PG1633+099, SA110, SA107 and PG1657+078) by the
Lulin One-meter Telescope operated by National Central University. The
summary of our observations and results are in Supplementary
Table~\ref{tab:rband_data}.

\begin{table}
\centering
\caption{\label{tab:rband_data}\textbf{ Log of R-band observations of \transient.}}
\begin{tabular}{llllll}
  \\
  \hline
  Start(UT)&	    End(UT)	&        Time(Mid Time) (UT)	&Filter&	Exp(s)&	Magnitude\\
  \hline
LOAO: \\
  2011-03-29T10:53:45& 2011-03-29T10:59:51&	2011-03-29T10:56:48&R&	300$\times$5& 22.45 $\pm$ 0.49	\\
  2011-03-30T10:50:24& 2011-03-30T11:47:47&	2011-03-30T11:19:06&	R&	300$\times$12&22.33 $\pm$ 0.22	\\
  2011-04-05T11:15:26& 2011-04-05T12:02:33&	2011-04-05T11:39:00&	R&	300$\times$6&	22.35 $\pm$ 0.25	\\
  2011-04-06T11:18:05& 2011-04-06T12:22:00&	2011-04-06T11:50:03&	R&	300$\times$12&22.20 $\pm$ 0.31	\\
  2011-04-08T10:09:28& 2011-04-08T11:29:07&	2011-04-08T10:49:17&
  R&	300$\times$14&22.73 $\pm$ 0.26	\\
\hline
Maidanak Obervatory: \\
 2011-04-12T22:49:39 &  2011-04-12T23:10:21 &      2011-04-12T22:59:59&  R&   $600 \times 2$  &   $ 22.49 \pm 0.12$     \\
  \hline
\end{tabular}
\end{table}

\paragraph{BOAO/KASINICS and TNG observations and data analysis}

We imaged the field of \transient\ with the Korea Astronomy and Space
science Institute (KASI) Near Infrared Camera System (KASINICS\cite{Moon08}) on
the 1.8m telescope at the Bohyunsan Optical Astronomy Observatory
(BOAO) in Korea, and with the NICS camera \cite{Baffa01} on the Italian 3.6m
Telescopio Nazionale Galileo (TNG), located in La Palma, Canary
Islands. Near-Infrared $J$, $H$ and $K/Ks$-band observations were carried out
on nights 30 and 31 March 2011 and 1, 3, 12, 13 and 21 April 2011. All
nights were clear, with seeing in the range $1.1'' - 1.3''$. The
complete observing log is reported in Supplementary Table~\ref{tab:nir_data}.  

Image reduction was carried out using the jitter pipeline data
reduction, part of the ESO-Eclipse package. Astrometry was performed
using the 2MASS (http://www.ipach.caltech.edu/2mass)
catalogue. Aperture photometry was made with the photom tool
implemented in the GAIA package and the photometric calibration to the
$Ks$ band was
done against the 2MASS catalogue. In order to minimize any systematic
effect, we performed differential photometry with respect to a
selection of local, isolated and unsaturated reference stars visible
in the field of view.

The \swift\ XRT light-curve (Supplementary Figure~\ref{fig:xrt_lc})
shows strong variability at essentially any epoch, and hints of an
analogous behaviour in the NIR band are seen when considering the whole
dataset. In particular, after the maximum NIR flux observed about 2
days after the first \swift\ trigger, global flux decreases are seen in
the NIR data (see Supplementary Figure~\ref{fig:nir_lc}).  About a
week after the trigger the NIR flux seems to have entered a
rather stable phase, which is still lasting more than three weeks
after the burst. It seems likely that the late-time NIR
data are due to the host galaxy contribution.

By means of almost simultaneous $z$, $J$, $H$, $K$, and $L$
observations (combining our data with those of Levan et al. \cite{Levan11}) in the
first days after the \swift\ trigger when the transient was brighter, we
can use the SED of the transient to evaluate the presence of substantial rest-frame
extinction.
Modelling the SED with a power-law in frequency we find that 
solutions with a rather flat or even increasing spectrum are favored
$(\nu^{\sim1/3})$. The required extinction, assuming a Milky Way extinction
curve\cite{Pei92}, is in the range $E_{B-V} \sim 1-3$, with a strong covariance with
the spectral index. The required extinction is indeed consistent with
the observed constancy of the R-band flux. The R-band emission is due to
the galaxy only and was already detected before the transient event\cite{Cenko11a},
while the transient contribution is depressed by local intrinsic
absorption well below the host-galaxy brightness.

\begin{figure}[t]
       \centering
        \parbox{6.5in}{
        \includegraphics[width=6in]{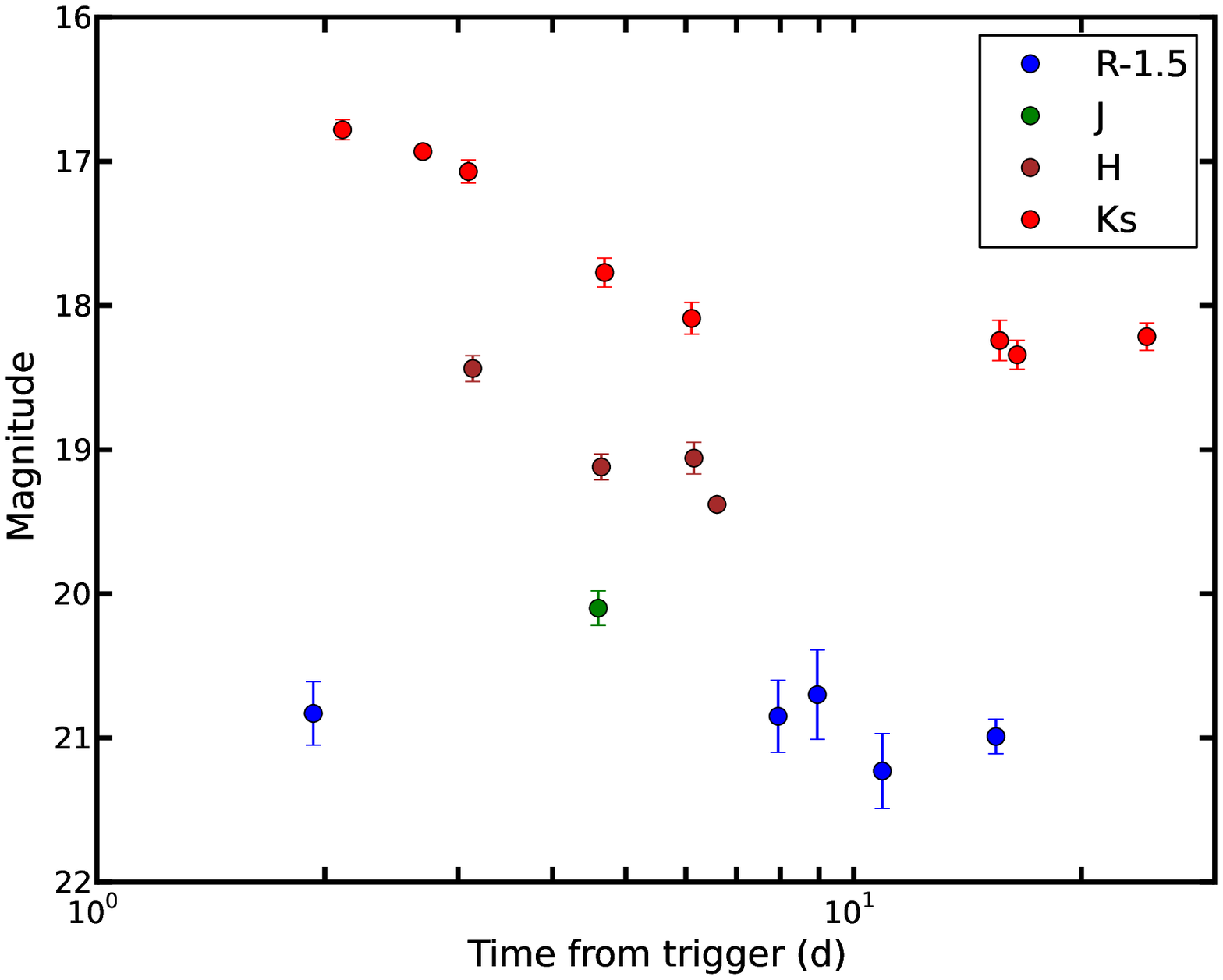}
        \centering\parbox{6in}{
          \caption{\label{fig:nir_lc}\textbf{Optical/NIR light curves in the R (blue), J
              (green), H (brown) and Ks (red) bands of \transient.} No
            variability has been detected in the R band while in the
            NIR a decay trend roughly consistent with X-ray data is
            visible. Error bars are 1$\sigma$.}}}

\end{figure}

\begin{table}
\centering
 \caption{\label{tab:nir_data}\textbf{Near Infrared data taken utilizing the TNG
      telescope and the BOAO telescope.}}
  \begin{tabular}{lllllll}
\\
    \hline
    Time of obs (UT)&  $\delta_T$ (days)&  Telescope&    Inst&      Filter&   Exp (s)&     Magnitude\\
    \hline
    2011-03-30.650910&   2.11081&	BOAO&	    KASINICS&	$K_s$	& 120$\times$43         & 16.78 $\pm$ 0.07	\\	
    2011-03-31.23164 &    2.69558&      TNG&         NICS &       $K$   &   20$\times$3$\times$40&    16.94 $\pm$ 0.04\\
    2011-03-31.636350&  3.09625&	BOAO&	    KASINICS&	$K_s$	&	90$\times$32   &17.07 $\pm$ 0.08\\
    2011-03-31.67763	&   3.13753&	BOAO&	    KASINICS&	$H$    & 60$\times$48	   &18.44 $\pm$ 0.09	\\
    2011-04-02.13674  &   4.59748&      TNG&         NICS &       $J$   &   60$\times$1$\times$40&    20.10 $\pm$ 0.12\\
    2011-04-02.17847  &   4.63973&      TNG&         NICS &       $H$   &   20$\times$3$\times$40&    19.12 $\pm$ 0.09\\
    2011-04-02.22289  &   4.68682&      TNG&         NICS &       $K$   &   20$\times$3$\times$40&    17.77 $\pm$ 0.10\\
    2011-04-03.648840	&   6.10874&	BOAO&	    KASINICS&	$K_s$    & 60$\times$64	   &18.09 $\pm$ 0.11\\
    2011-04-03.689630	&   6.14953&	BOAO&	    KASINICS&	$H$
    & 60$\times$96	   &19.06 $\pm$ 0.11\\
    2011-04-13.12765  &  15.59158&      TNG&         NICS &       $K$  &    20$\times$3$\times$20&    18.24 $\pm$ 0.14\\
    2011-04-13.98145  &  16.44759&     TNG &        NICS  &      $K$   &   20$\times$3$\times$40 &   18.34 $\pm$ 0.10\\
    2011-04-21.927812 & 24.40270& TNG & NICS & $K$ & 20$\times$3$\times$30 & 18.22 $\pm$ 0.10 \\
    \hline
  \end{tabular}
\end{table}

\subsection*{Historical X-ray/$\gamma$-ray Upper Limits}

We have examined archival databases from several X-ray and $\gamma$-ray instruments
that have observed the location of \transient, either through wide field
surveying or in serendipitous pointings, to search for earlier
detections of a source from this direction at a variety of
wavelengths.

\paragraph{BAT Historical upper limits}

The BAT instrument monitors about 80\% of the sky daily. Examining the
BAT Survey data at the location of \transient, we find that the hard
X-ray source never exceeded the $6\sigma$ significance level when it
was in the BAT field of view between February 2005 and March 2011 
(count rate and flux limits vary with each observation), and was not
found in any of the BAT survey studies\cite{Tueller10,Cusumano10} down
to a limiting flux of $10^{-11}$ \cgs\ (15-150 keV), about two orders
of magnitude below the average flux for the first 3 days after the BAT
trigger.

The BAT data are also monitored daily for transient events.  We
searched the entire BAT data set from 12 February 2005 to 28 March  2011,
searching for previous detections of \transient\ on three timescales.
The source is not detected on 16-day timescales, with a $3\sigma$
upper limit of $0.0011$ \batrate (about 5 mCrab).  On one 1-day
timescales, we obtain marginal detections exceeding $3\sigma$ on 31
March 2009, 14 September 2009, and 14 March 2011, all at about $0.0036
\pm 0.0011$ \batrate; this rate is consistent with the expected rate
of false $3\sigma$ fluctuations in a search of 1866 independent
measurements, and we therefore do not consider these detections to be
significant.  With the exception of these three days, the 1-day
$3\sigma$ upper limits vary depending on the total exposure of the
source, but are between $0.0010$ and $0.0015$ \batrate\ for 90\% of
the observations.  On shorter timescales, we can state that the source
never exceeded the $6\sigma$ significance level on any timescale
between 64\,s and 1200\,s when it was in the BAT field of view (count
rate and flux limits vary with each observation) until it triggered
the BAT in a 1200 second image trigger on 28 March 2011.

\paragraph{\rosat\ Historical Upper Limits}

\transient\ was serendipitously observed in a 6.5\,ks \rosat\ PSPC-B
observation made on 3 April 1992 covering the energy range $0.1 -
2.35$ keV.  The source is $\sim 42'$ off axis and is affected by
shadowing from one of the PSPC windows support ribs, giving an
effective exposure time of 3.3 ks.  The $3\sigma$ upper limit\cite{Kraft91} for the PSPC
count rate is $8.3 \times 10^{-3}$ cts s$^{-1}$, corresponding to an observed
$0.3-10$ keV flux limit of $< 2.9 \times 10^{-13}$ \cgs\ (observed) for any
of the spectral models that fit the XRT data.  We note that no source
is detected at even the $1.5\sigma$ level.  
This limit is more than an order of magnitude below the faintest
portion of the XRT light curve, and four orders of magnitude below the
brightest flares.
Upper limits on the flux for different
spectral models for \transient\ are given in Supplementary
Table~\ref{tab:rosat_upper}.

The \rosat\ All-Sky Survey also covered this field with a relatively
long exposure due to its high ecliptic latitude, collecting 940\,s of data between 11 July 1990 and 13
August 1991.  These data provide a 3$\sigma$ PSPC count rate upper
limit\cite{Kraft91} of $7.3 \times 10^{-3}$ cts s$^{-1}$, with a corresponding 0.3--10
keV flux limit of $< 2.5 \times 10^{-13}$ ergs cm$^{-2}$ s$^{-1}$.

\begin{table}
\centering
\caption{\label{tab:rosat_upper}\textbf{$3\sigma$ upper limits based on \rosat\ data for various spectral models}}
\begin{tabular}{lcccc}
\\
\hline \\
& \multicolumn{4}{c}{All fluxes are 3$\sigma$ upper limits, \cgs
  (0.3--10 keV)} \\
& \\
& \multicolumn{2}{c}{3 April 1992 Pointed Observation} &
\multicolumn{2}{c}{RASS Observation} \\
Model \qquad \qquad & Observed Flux & Unabsorbed Flux & Observed Flux & Unabsorbed Flux \\
\hline \\
\multicolumn{5}{l}{Simple absorbed power law model ($N_{\rm H} =
  7.73\times10^{21} $ cm$^{-2}, \Gamma = 2.80$):} \\
& \\
& 	$< 2.88\times 10^{-13}$
& 	$< 1.21\times 10^{-12}$  \qquad \qquad
& 	$< 2.52\times 10^{-13}$
& 	$< 1.06\times 10^{-12}$ \\
\hline \\
\multicolumn{5}{l}{Broken power law model ($N_{\rm H} = 1.30 \times 10^{22} $ cm$^{-2},
\Gamma_1 = 4.98, E_{break} = 1.96$ keV, $\Gamma_2 = 2.48$):} \\
& \\
&	$< 2.88\times 10^{-13}$ 
&	$< 1.83\times 10^{-11}$ 
&	$< 2.52\times 10^{-13}$ 
&	$< 1.59\times 10^{-11}$ \\
\hline \\
\multicolumn{5}{l}{Disk black body + power law model ($N_{\rm H} = 1.24\times10^{22}$ cm$^{-2}, T_{\rm
    in} = 175 {\rm eV}, \Gamma = 2.40$):} \\
 & \\
 &	$< 2.90\times 10^{-13}$ 
 &	$< 4.24\times 10^{-12}$
 \\
\hline
\end{tabular}
\end{table}

\paragraph{\xmm\ Historical Upper Limits}

Although there were no pointed observations of the \transient\ field
by \xmm\ prior to the outburst, between 26 August 2001 and 25
March 2011 \xmm\ slewed over this object on one occasion at 01:42:32
UT on 27 July 2005. The EPIC-pn routinely takes data during such
slews\cite{Saxton08}.
No detection was made during the 3.9\,s exposure made with the medium filter,
resulting in a $2\sigma$ upper limit of $<0.79$ cts s$^{-1}$ (0.2--12 keV),
calculated following the Bayesian approach of Kraft, Burrows \&
Nousek\cite{Kraft91}.  Unabsorbed 0.2$-$12~keV flux limits were derived for
the following three spectral models: a power law spectrum of photon index
$\Gamma=2.8$ absorbed by a Hydrogen equivalent column density of $N_H = 7.7
\times 10^{21}$~cm$^{-2}$, a broken power law with $\Gamma_1 = 4.98,
\Gamma_2 = 2.48, E_{break} = 1.96$~keV, and $N_H =1.30 \times
10^{22}$~cm$^{-2}$, and a disk blackbody plus power law model with $T_{in}
= 175$~eV, $\Gamma = 2.40$, and $N_H = 1.24 \times 10^{22}$~cm$^{-2}$;
these flux limits were $< 2.9 \times 10^{-12}$ \cgs, $< 3.2 \times
10^{-12}$ \cgs, and $< 3.2 \times 10^{-12}$ \cgs, respectively.

\paragraph{\textit{MAXI} Historical Upper Limits}

To investigate whether \textit{MAXI} detected the \transient\ before the current
outburst, we have analysed historical \textit{MAXI} GSC data for the period of 17
August 2009 to 1 March 2011 (see Supplementary
Figure~\ref{fig:maxi_longterm_lc}).  No detection of \transient\ is
found in those data, giving a 90\% confidence level upper limit of
$\sim 1.1 \times 10^{-11}$ \cgs\ (2-20 keV). In addition to this,
utilizing data that has had additional cleaning applied, but covers a
narrower time period ( between 1 September 2009 and 31 March 2010), we
obtain a deeper $90\%$ confidence level upper limit of $2.7 \times
10^{-12}$ \cgs\ (4-20 keV).

\begin{figure}[t]
       \centering
        \parbox{6.5in}{
        \includegraphics[width=6in]{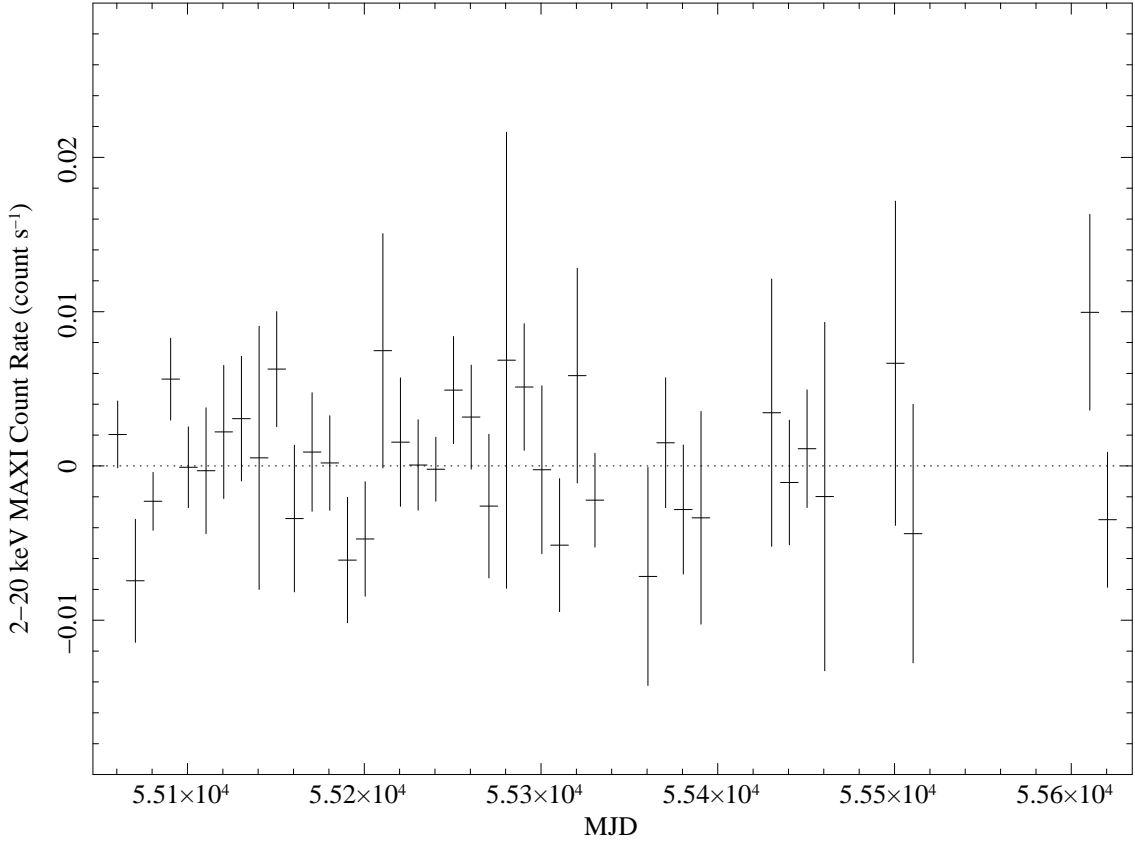}
        \centering\parbox{6in}{
\caption{\label{fig:maxi_longterm_lc}\textbf{\textit{MAXI} light curve of \transient\ for the period August 17
    2009 to 1 March 2011.} Note that the count rates seen in this light
    curve are consistent with a non-detection of \transient\ in this period}
}}
\end{figure}

\paragraph{\fermi\ Historical Upper Limits}

The analysis of the full 32-month \fermi/LAT dataset does not reveal any new
source within 3 degrees of the position of the transient. This places
95\% confidence upper
limits of $1.7 \times 10^{-8}$ \latrate\ (100 MeV-10 GeV) and $1.5 \times
10^{-10}$ \latrate\ (1 GeV - 300 GeV) on any persistent prior
emission. Given the variable nature of this source, a search for emission
was performed on shorter timescales of 2 days and 5 days over the
\fermi\ mission lifetime. No significant variation from the mean background level
is observed in the light curve (see Supplementary Figure~\ref{fig:fermi_ul_lc}).

\begin{figure}[t]
       \centering
        \parbox{6.5in}{
        \includegraphics[width=6in]{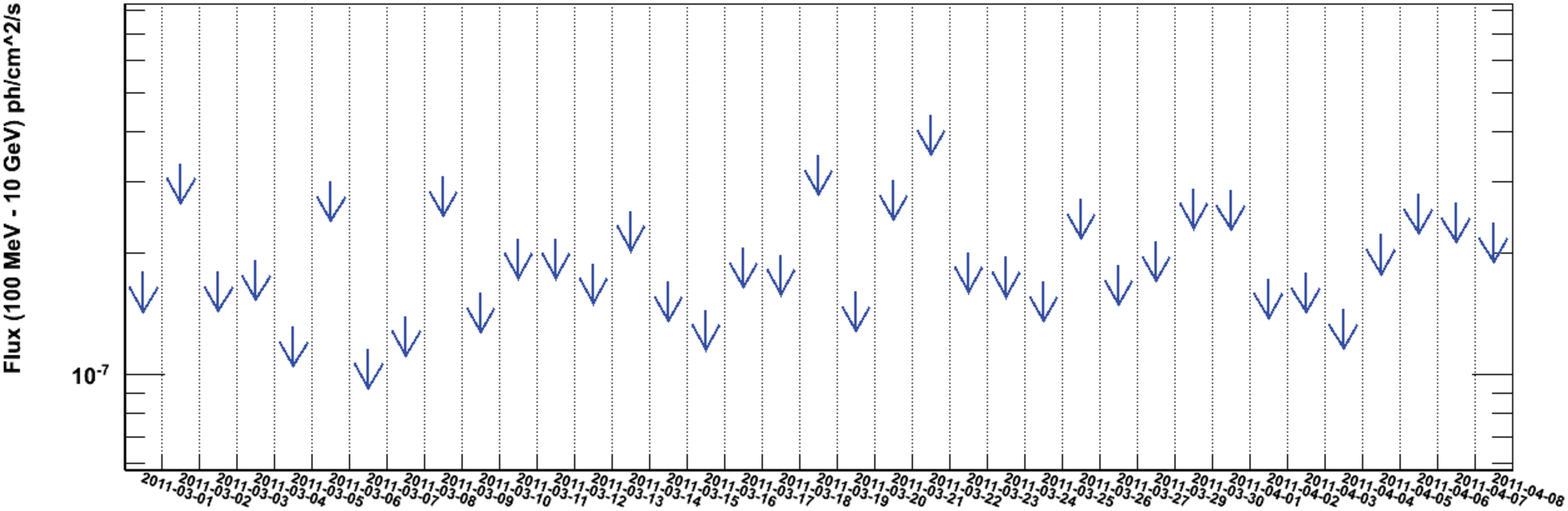}
        \centering\parbox{6in}{
\caption{\label{fig:fermi_ul_lc}\textbf{\fermi\ 95\% confidence upper limit light curve of \transient\ for the period August 17
    2009 to 1 March 2011.}}
}}
\end{figure}

\section*{Supplementary Discussion}

\subsection*{Constraints on the Mass of the Central Black Hole}

An upper limit to the mass of the central black hole in the galaxy
where the transient was observed can be obtained through the black
hole mass $vs$ host bulge luminosity relation \cite{Marconi03}.  Using
the measured $B$ and $H$ magnitudes of the host galaxy from published
reports \cite{Leloudas11, Fruchter11} and from our own analysis of
{\it Hubble Space Telescope} WFC3 observations, and taking a luminosity
distance of of $d_{\rm L}=1.88\;$Gpc, we infer the following
luminosities relative to the Sun: $\log (L_{\rm B}/L_{\rm \odot, B}) =
9.20$ and $\log (L_{\rm H}/L_{\rm \odot, H}) = 9.58$ (the $B$-band
luminosity includes a Galactic extinction correction of $A_{\rm
  B}=0.08\;$mag). These luminosity estimates lead to a black hole mass
of $M_{\rm bh}\approx 2\times 10^7\;{\rm M}_\odot$, with a systematic
uncertainty of a factor of 2 resulting from the scatter in the
empirical relation between the host luminosity and black hole
mass. Moreover, this empirical formula relates the luminosity of the
{\it spheroid} of the host galaxy (i.e, its bulge) to the black hole
mass. However, the spectrum of the host galaxy shows the
spectroscopic signature of star formation \cite{Cenko11b}, which
suggests that it is a spiral.  Since the magnitudes we have used refer
to the entire galaxy, not just its bulge, the above value is an upper
limit to the black hole mass.

An independent constraint on the black hole mass can be obtained from
the minimum X--ray variability time scale, which is observed to be
$\delta t_{\rm min} \sim 100$\,s (3$\sigma$ value). 
This time scale
constrains the size of the central engine since, for Schwarzschild black
hole, the minimum variability time scale is in its rest frame is
$\delta t_{\rm min} \sim r_{\rm S}/c \sim 100\;M_6~{\rm s}$, where $M_6 =
M_{\rm bh}/10^6 M_\odot$.  Thus,
\begin{equation}
M_{\rm bh} \sim 7.4\times 10^6 
\left(\frac{\delta t_{\rm obs}}{100\;{\rm s}}\right)~M_\odot~.
\end{equation}
Much larger or much smaller masses are unlikely for the following
reasons. For larger masses, a variability time scale much shorter than
the above constraint was never observed in the X-ray band in other
systems containing supermassive black holes (e.g. AGNs)\footnote{More
  rapid TeV variability has been observed in some blazars, which may
  be attributed to additional local Lorentz boost within a jet
  (e.g. jet in the jet) \cite{giannios09}.}.  For much smaller masses
(e.g. stellar--mass black holes), one would expect to see much shorter
variability time scales in the light curves.

Some nearby low--luminosity GRBs with a stellar--mass central engine
(e.g. GRB\,060218) do have a smooth light curve without a noticeable fast
variability component. However, GRB\,060218 has only one peak.  It would be
contrived to have a smaller black hole smear up all the shorter
variability time scales but only keep the 100\,s time scale. We
conclude that the black hole mass is
likely $M_{\rm bh}\sim (10^6 - 2 \times 10^7) M_\odot$ ($1 < M_6 < 20$).

\subsection*{Energetics and Their Implications}

After the initial flare, the source settled down to an 
X--ray flux of
$F_{\rm X} \sim 10^{-9}~{\rm erg~cm^{-2}~s^{-1}}$. 
The X--ray to $\gamma$--ray luminosity is 
\begin{equation}
L_{\rm \gamma,X}=4.2 \times 10^{46}~{\rm erg~s^{-1}}
\left(\frac{F_{\rm \gamma,X}} {10^{-10}~{\rm erg~cm^{-2}~s^{-1}}}\right)\, ,
\end{equation}
assuming the above luminosity distance. 
This luminosity is well above
the Eddington limit for a $10^6\; {\rm M}_\odot$ black hole 
($L_{\rm Edd} = 1.3 \times 10^{44}\, M_6\; {\rm erg~s^{-1}}$ 
with $M_{\rm bh}= 10^6\, M_6\; {\rm M}_\odot$), suggesting that the emission
originates from a relativistic jet rather than an accretion disk.

The above luminosity, along with the coincidence of the event with the
center of the host galaxy\cite{Fruchter11}, lead us to consider the
onset of accretion--powered nuclear activity in the host galaxy. 
This could be fueled either by the tidal disruption of a star by the
central black \cite{Rees88} hole or by the supply of gas from
another source.  

To estimate the mass accretion rate and the total mass accreted, 
let us assume that the accretion rate is
$\dot M = f \dot M_{\rm Edd}$, where $\dot M_{\rm Edd}= L_{\rm Edd}/(\eta c^2)$,
and $\eta$ is the usual accretion efficiency.
We should allow the factor $f$ to be larger than unity, since
the accretion rate, especially for the early phases of the tidal disruption event,
can be super--Eddington.
Therefore
\begin{equation}
\dot M = f \dot M_{\rm Edd} = f\, {L_{\rm Edd} \over \eta c^2} = 
1.4\times 10^{23} M_6 {f \over \eta} \,\,{\rm g\,\,  s^{-1}}
\end{equation}
giving a total mass, accreted until now, of the order
\begin{equation}
M = \Delta t \dot M \sim 2\times 10^{-4}  {f \over \eta} \,\, {\Delta t \over 1\, {\rm month}}  \,  M_\odot
\end{equation}
leading to a few per cent of a solar mass if $f/\eta\sim 10^2$.
In fact, the measured radiative energy in the 0.4--13.5~keV rest frame
band for the
first 23.4 days of activity, $3 \times 10^{53}$~erg, corresponds to
$0.016 (\eta/0.1) M_\odot$, assuming isotropic emission.

If the source of fuel is the tidal disruption of a star by the black
hole, then we  expect an initially super--Eddington flare from the
accretion of bound debris, followed by a steady decline that traces
the rate of return of post-disruption debris, $\left(\dot M\propto
(t-t_0)^{-5/3}\right)$ \cite{Phinney89, Evans89, Lodato09}. The monotonic
decline phase could lag the initial flare by as much as 1--2 months
\cite{Lodato09}, though the X-ray light curve (Figure 1 in the Letter)
suggests that the decline phase may have begun within a few days of
the flare onset for \transient. Therefore, long--term monitoring of the source in
search of this specific rate of decline will be able to test the tidal
disruption scenario.

\subsection*{Modelling the Spectral Energy Distribution of the Emerging Jet}

\subsubsection*{Available Data}

\begin{description}

\item[\it Radio Observations --] 
The 8.4~GHz flux density was observed on 1
  April\cite{Bower11}. The 13.5~GHz, 15~GHz and
  100~GHz (3~mm) fluxes were also reported in GCN Circulars \cite{Pooley11, Tirado11,
    Mooley11}. Although important, these data will not be used to 
    for the modelling of the X--ray and IR emitting region, since 
    the radio emission must come from a larger region, not to be
    self--absorbed.
Additional radio observations and their interpretation are treated by
Zauderer et al.\cite{Zauderer11}.
  
\item [\it IR and Optical Observations --] 
In the Supplementary Methods we present the near--infrared observations performed with 
the TNG and BOAO telescopes. 
Of these data, we have selected for the modelling those corresponding to the 
peak of the IR light curve (approximately 2 days after trigger),
at the first minimum of the X-ray light curve ($\sim$4.5 days), and
at an intermediate state about a week after the trigger.
The large column density derived from the X-ray fitting,
of the order of $N_{\rm H}\sim 10^{22}$ cm$^{-2}$, suggests
a large value of the extinction, even if the dust to gas
ratio is not known.
We de-redden our data with a rest frame $A_V=4.5$, as discussed in the
Supplementary Methods,
approximately as in Bloom et al.\cite{Bloom11}.

\item[\it X-Ray Observations --] 
We have chosen to show the spectra for the peak (31 hours from trigger), including BAT data, 
for the minimum after 4.5 days, and for the intermediate persistent 
flux state that began a few days after the BAT trigger.

\item [\it $\gamma$-ray Observations --] We use the upper limit in
  the {\it Fermi}/LAT band reported in the Supplementary Methods.

\item [\it TeV Observations --] 

  The {\it VERITAS} Collaboration \cite{Aliu11} set an upper limit on VHE
  emission during the early ``flaring" period March 29--March 31 and
  the later ``quiescent" period, April 1--April 15.  The two upper
  limits are reported in the SED shown in Supplementary
  Figure~\ref{F:sed-one}--\ref{F:sed-three}.

\end{description}

\subsection*{Assumptions and Observational Constraints on the Model}

We assume that the spectrum is produced by the synchrotron and inverse 
Compton (IC) mechanisms.  
In general, the seed photons for the inverse Compton process can be produced by the synchrotron 
process itself (Synchrotron Self-Compton, SSC) or can be produced outside the jet 
(External Compton, EC).
A new-born blazar has had no time to build up a broad line region (i.e., to populate the
surroundings of the jet with external photons), but some contribution to the EC
process can come from photons produced in the accretion disk.
The following is a summary of information we have used to guide the
development of the model:

\begin{itemize} 

\item
a limit on the size of the X-ray emitting region from the variability
timescale, $t_{\rm v}\sim 100$ s;

\item the isotropic luminosity $L_{\rm obs}$; at the peak of the flare
  is $L_{\rm obs}= 10^{48}L_{\rm syn, 48}$ erg s$^{-1}$; there is
  variability by a factor $>100$;

\item some hints on the peak energy: {\it Fermi} and {\it VERITAS} did not detect the
  source;

\item the slope of the line connecting the IR to the X-rays is hard,
  possibly as hard as $\nu^{1/3}$;

\item it is the first time we have ever observed a flare of this kind, therefore
  the phenomenon is relatively rare.

\item the upper limit on the black hole mass derived above is $2\times
  10^7\;{\rm M}_\odot$ and a mass compatible with the observed
  variability is $\sim 7 \times 10^6\;{\rm M}_\odot$.

\end{itemize}

For blazar jets, one can find a unique solution for modelling their
spectral energy distribution (SED) when the emission is SSC, and when
we know the peak frequencies of the synchrotron and Compton spectra,
the slopes before and after the peaks, the peak flux levels, and the
variability timescale.  Here we do not know anything about the Compton
component, and we can only guess the peak frequency (and flux) of the
synchrotron component.  Therefore we need additional assumptions in
order to find a reasonable fit.  

One possibility is to assume that: ($i$) the
source is magnetically dominated (note that this is {\it not} the case
for blazars, whose jet powers are dominated by the kinetic energy of
particles), and ($ii$) the self Compton flux must be absorbed by the
$\gamma$--$\gamma \to e^\pm$ process.  The first assumption comes from
the very hard optical to X-ray spectrum, requiring a low energy
cut-off in the particle distribution, and thus indicates a
particle-starved source, 
while the second requirement comes from the {\it Fermi} and {\it VERITAS} upper limits.

\begin{description}

\item[\it The size --] We assume for simplicity a spherical source of
  radius $R$ at some distance $d$ from the black hole.  We define $\mathcal{D}\equiv
  1/[\Gamma(1-\beta\cos\psi_{\rm v})]$ as the Doppler factor, where
  $\psi_{\rm v}$ is the viewing angle.  Note that for blazars one
  assumes a mono-directional velocity (unlike in GRBs).  The size is
  constrained by:
\begin{equation}
R \le ct_{\rm v} {\mathcal{D} \over 1+z}
\label{Q:R}
\end{equation}

\item[\it Magnetic field --] We assume that the bulk of the luminosity
  is produced by synchrotron radiation.  We further assume that the jet is
  magnetically dominated.  This is because of the very hard slope
  between the optical and the X-rays, implying that there are no low
  energy electrons in the source.  If there were, then they would
  severely overproduce the optical flux.  One possibility is that the
  electrons are continuously reaccelerated/heated, in such a way that
  there are no electrons below, say, $\gamma\sim10^2-10^3$.  So we need a
  ``thermal bath", but with relatively few electrons inside.
  Otherwise the mean energy of the particles will be small.  As a
  consequence, the jet is particle-starved, and the jet power must be
  carried by the magnetic field.  The jet is however capable of
  producing a lot of radiation: the isotropic luminosity we see is $L_{\rm obs} \sim
  10^{48}$ erg s$^{-1}$, but this is beamed.  The real power spent by
  the jet to produce $L_{\rm obs}$ is
\begin{equation}
P_{\rm r}\sim {L_{\rm obs}\over \Gamma^2}
\end{equation}
up to factor of order unity (see e.g. Ghisellini et al. 2010, MNRAS,
402, 497).  If the jet is particle-starved, this power must be given
by the Poynting flux, and therefore we require:
\begin{equation}
P_{\rm B} \, =\, \pi R^2 \Gamma^2 c {B^2 \over 8\pi} \, >\, {L_{\rm
    obs}\over \Gamma^2}
\end{equation}
Together with Supplementary Equation~\ref{Q:R} this gives the limit:
\begin{equation}
B\, > \, {1+z\over \Gamma^2 \mathcal{D} t_{\rm v}}\,  \left( {L_{\rm obs} \over 8 c^3}\right)^{1/2}
\label{Q:B}
\end{equation}

\item[\it Pair production --] Since we are dealing with a compact
  source, with a large produced synchrotron luminosity, it is
  conceivable that the self-Compton emission would be relevant.  But
  the upper limits by {\it Fermi} tell us that the observed
  $\gamma$-ray flux is much less than the synchrotron flux.  One way
  to account for that is pair opacity.  We may therefore require that
  the optical depth for pair production is larger than unity:
\begin{equation}
\tau_{\gamma\gamma} \sim {\sigma_{\rm T}\over 5} \, 
{ U^\prime_{\rm ssc} \over m_{\rm e} c^2 } R >1
\end{equation}
where $U^\prime_{\rm ssc}$ is the comoving synchrotron self-Compton
radiation energy density, and $U^\prime_{\rm ssc}/m_{\rm e}c^2$ is
(very approximately!) the number density of photons at threshold,
where $\sigma_{\gamma\gamma}\sim \sigma_{\rm T}/5$.  We then have:
\begin{equation}
{\sigma_{\rm T}\over 5} \, { L^\prime_{\rm ssc} \over 4 \pi R^2c  m_{\rm e} c^2 } R 
\, \, =\,\,  {\sigma_{\rm T}\over 20\pi } \, 
{ L^\prime_{\rm ssc}\over L^\prime_{\rm syn}}{ ( L_{\rm syn}/\mathcal{D}^4) \over R   m_{\rm e} c^3 } 
\, \, =\, \,  {\sigma_{\rm T}\over 20\pi } \, 
{ L^\prime_{\rm ssc}\over L^\prime_{\rm syn}}{L_{\rm syn} (1+z) \over \mathcal{D}^5 ct_{\rm v}   
m_{\rm e} c^3 }  >1
\label{Q:pairs}
\end{equation}

\item[\it Distance from the black hole --] If we assume a conical jet
  of semi-aperture angle $\theta \equiv 0.1 \theta_{-1}$ we have
\begin{equation}
d \, \sim \, {R\over \theta} \, <\, {10 c t_{\rm v} \mathcal{D} \over \theta_{\rm -1} (1+z)} 
\end{equation}
Since we continue not to detect the source in $\gamma$-rays, while
the X-rays vary widely, it is likely that we are observing some
sort of a standing shock, or (magnetically dominated) internal
shocks.  The emission cannot be due to a single traveling blob.  A
single blob would travel rapidly, expand, and die in a short time.

\item[\it Some numbers --]
From Supplementary Equation~\ref{Q:pairs} we can set a limit on $\mathcal{D}$:
\begin{equation}
\mathcal{D} \, <\, 11.4 \left( {L_{\rm syn, 48} \over t_{\rm v, 2}}\, 
{L^\prime_{\rm ssc}\over L^\prime_{\rm syn}} \right)^{1/5}
\end{equation}
where $t_{\rm v} \equiv 10^2 t_{\rm v, 2}$ s. 
Using this limit in Supplementary Equation~\ref{Q:B}, we derive a lower limit on $B$:
\begin{equation}
B \, >\, 620\,\, 
{ (\mathcal{D}/\Gamma)^3  \over L_{\rm syn, 48}^{1/10} t_{\rm v, 2}^{2/5}} \,
\left( {L^\prime_{\rm syn} \over L^\prime_{\rm ssc} }   \right)^{3/5} \,\, {\rm Gauss}
\end{equation}
The size should be:
\begin{equation}
R\,\, <\,\, 2.5\times 10^{13} t_{\rm v,2}^{4/5} \, L_{\rm ssc, 48}^{1/5}\,\,\, {\rm cm}
\end{equation}
The distance from the black hole, in units of the Schwarzschild radius is
\begin{equation}
{d \over R_{\rm S}} \, <\, 833 \, 
{t_{\rm v,2}^{4/5} \, L_{\rm ssc, 48}^{1/5} \over \theta_{-1} M_6}
\end{equation}

\end{description}

\subsection*{Modelling Procedure and Related Considerations}

We use the model described in detail in Ghisellini et al.\cite{Ghisellini09}.  The
emitting region is assumed spherical, of size $R_{\rm blob}$, moving
with a bulk Lorentz factor $\Gamma$ and located at a distance $d$
from the black hole of mass $M$.  The bolometric luminosity of the
accretion disk is $L_{\rm disk}$.

The particle energy distribution $N(\gamma)$ [cm$^{-3}$] is calculated
by solving the continuity equation where particle injection, radiative
cooling and pair production (via the $\gamma$--$\gamma \to e^\pm$
process) are taken into account.  The created pairs contribute to the
emission.  The injection function $Q(\gamma)$ [cm$^{-3}$ s$^{-1}$] is
assumed to be a smoothly joined broken power-law, with a slope
$Q(\gamma)\propto \gamma^{-{s_1}}$ and $\gamma^{-{s_2}}$ below and
above a break energy $\gamma_{\rm b}$:
\begin{equation}
Q(\gamma)  \, = \, Q_0\, { (\gamma/\gamma_{\rm b})^{-s_1} \over 1+
(\gamma/\gamma_{\rm b})^{-s_1+s_2} }
\label{qgamma}
\end{equation}
In the specific application here, we assumed that electrons below a
given $\gamma_0$ simply disappear.

The total power injected into the source in the form of relativistic
electrons is $P^\prime_{\rm i}=m_{\rm e}c^2 V\int Q(\gamma)\gamma
d\gamma$, where $V=(4\pi/3)R_{\rm blob}^3$ is the volume of the
emitting region.

The injection process lasts for a light crossing time $R_{\rm
blob}/c$, and we calculate $N(\gamma)$ at this time.  This
assumption comes from the fact that even if injection lasted longer,
adiabatic losses caused by the expansion of the source (which is
travelling while emitting) and the corresponding decrease of the
magnetic field would make the observed flux decrease.  Therefore
the computed spectra correspond to the maximum of a flaring episode.

Above and below the inner parts of the accretion disk there is an
X-ray emitting corona of luminosity $L_{\rm X}$ (it is fixed at a
level of 30\% of $L_{\rm d}$).  Its spectrum is a power law of energy
index $\alpha_X$ ending with a exponential cutoff at $E_{\rm c}=$150\,keV.  
The specific energy density (i.e., as a function of frequency) of
the disk and the corona are calculated in the comoving frame of the
emitting blob, and used to properly calculate the resulting external
inverse Compton spectrum.  The internally produced synchrotron
emission is used to calculate the synchrotron self Compton (SSC) flux.
In this specific case we assume no Broad Line Region; the disk and the
coronal radiation are negligible if the emitting region is at a large
distance from the disk, but  they become important producers 
of seed photons if the distance $d$ of the emitting
blob from the disk is of the same order of (or less than) the outer radius of the disk.

\subsection*{Results and Discussion}

Supplementary Table~\ref{T:para} lists three sets of parameters adopted to generate
three different models. In Supplementary Figures~\ref{F:sed-one}--\ref{F:sed-three}
we show the models resulting from these input parameters.
Supplementary Table
\ref{T:powers} lists the power carried by the jet in the form of
radiation ($P_{\rm r}$), magnetic field ($P_{\rm B}$), emitting
electrons ($P_{\rm e}$, no cold electron component is assumed) and
cold protons ($P_{\rm p}$, assuming one proton per emitting electron).
All the powers are calculated as
\begin{equation}
P_i  \, =\, \pi R_{\rm blob}^2 \Gamma^2\beta c \, U^\prime_i
\end{equation}
where $U^\prime_i$ is the energy density of the $i$ component, as
measured in the comoving frame.  We note the following regarding
Supplementary Table
\ref{T:powers}:

\begin{itemize}

\item
The power carried in the form of radiation, $P_{\rm r} =\pi R_{\rm
  blob}^2 \Gamma^2\beta c \, U^\prime_{\rm rad}$, can be rewritten
as [using $U^\prime_{\rm rad}=L^\prime/(4\pi R_{\rm blob}^2 c)$]:
\begin{equation}
P_{\rm r}  \, =\,  L^\prime {\Gamma^2 \over 4} \, =\, L {\Gamma^2 \over 4 \mathcal{D}^4}
\, \sim \, L {1 \over 4 \mathcal{D}^2}
\end{equation} 
where $L$ is the total observed non-thermal luminosity ($L^\prime$ is
in the comoving frame) and $U^\prime_{\rm rad}$ is the radiation
energy density produced by the jet (i.e.  excluding the external
components).  The last equality assumes $\psi_{\rm v}\sim 1/\Gamma$.

\item
When calculating $P_{\rm e}$ (the jet power in bulk motion of emitting
electrons) we include their average energy, i.e.  $U^\prime_{\rm e}=
n_{\rm e} \langle\gamma\rangle m_{\rm e} c^2$.

\item
For $P_{\rm p}$ (the jet power in bulk motion of cold protons) we have
assumed that there is one proton per emitting electron,
i.e. electron-positron pairs are negligible.  This is unimportant
for the models shown in Supplementary Figure~\ref{F:sed-one} and~\ref{F:sed-two},
since in these cases the mean energy of the electrons is comparable to the rest
mass of protons ($\langle\gamma\rangle m_{\rm e} \approx m_{\rm p}$).
However, for the model shown in Supplementary Figure~\ref{F:sed-three}, $P_{\rm p}$ is the 
dominant form of power.

\item
$P_{\rm B}$ is derived using the magnetic field found from the model
fitting.

\end{itemize}

In summary, we have studied the extreme cases of a jet whose power is largely dominated
by the magnetic field (Models 1,2; Supplementary
Figures~\ref{F:sed-one} and \ref{F:sed-two})
or by the kinetic energy of the matter (Model 3; Supplementary Figure~\ref{F:sed-three}).

\subsubsection*{A magnetic field-dominated jet}
Our rationale for this model is that since the
  accretion can be super-Eddington, the density of the accreting
  matter is large and can sustain a very large $B$-field, which
  launches the jet.  In the dissipation region, the $B$-field is still
  large, possibly because it has not completed the acceleration of the
  matter.  Dissipation can occur at the expense of the magnetic field,
  through reconnection. 

Models shown in Supplementary Figures~\ref{F:sed-one} and \ref{F:sed-two}
rely on the SSC process (the magnetic energy density 
dominates over the radiation energy density of external seed photons), 
but adopt different black hole masses: $10^6\;{\rm M}_\odot$ and 
$10^7\;{\rm M}_\odot$, respectively. 
In the first model the disk luminosity is kept constant at the Eddington value,
even if the jet luminosity and its power change, 
while in the second model we have assumed that the jet luminosity is 
tracking a fastly varying accretion luminosity.
In this case, for the high state we have assumed a super-Eddington luminosity,
and nearly Eddington for the low state.
As can be seen, the disk luminosity can barely contribute to the soft X-ray flux.

The dotted orange line in both models shows the flux emitted from a much larger
region of the jet, producing the radio flux.

\subsubsection*{A matter dominated jet}

For this model we assume that even if, at the start, 
the jet is dominated by the magnetic field, nevertheless the Poynting flux 
is able to accelerate the jet to its final $\Gamma$ before the
dissipation region. 
The jet is then dominated by the kinetic energy of matter, therefore 
dissipation is at the expense of the latter, as in all other blazars.
The source of the seed photons is
an accretion disk with an outer radius of $\gtrsim 300$ Schwarzschild
radii (it is 500 $R_{\rm S}$ for the model shown in Supplementary Figure~\ref{F:sed-three}).
This model is similar to models used for blazars, but with a
particle energy distribution that does not extend up to such large energies 
(i.e. $\gamma_{\rm max}\sim$80--100),
with most of the electrons at low energies.
Therefore in this case we have a sort of bulk-Compton process, in which
relatively cold electrons scatter the seed photons coming from the outer
radius of the disk.
This model is similar to the one presented by Bloom et al.\cite{Bloom11},
but in our case the IR and X-ray photons are produced in the same region. 
The large $L_X/L_{\rm IR}$ ratio is due to the large ratio between
the energy densities of the external radiation and the magnetic field.
This requires a small magnetic field, and therefore a matter-dominated jet.

{\small
\begin{table} 
\centering
\caption{\textbf{Input parameters used to model the SED.}
Col. [1]: state;
Col. [2]: size of the emitting region, in units of $\mathsf{10^{15}}$ cm;
Col. [3]: black hole mass in solar masses;
Col. [4]: Disk luminosity in units of $\mathsf{10^{45}}$ erg s$\mathsf{^{-1}}$ and ($L_{\rm d}/L_{\rm Edd}$);
Col. [5]: power injected in the blob calculated in the comoving frame, in units of $\mathsf{10^{45}}$ erg s$\mathsf{^{-1}}$; 
Col. [6]: magnetic field in Gauss;
Col. [7]: bulk Lorentz factor;
Col. [8]: viewing angle in degrees;
Col. [9]: Doppler factor;
Col. [10], [11] and [12]: minimum, break and maximum random Lorentz factors of the injected electrons;
Col. [13] and [14]: slopes of the injected electron distribution [$Q(\gamma)$] below and above $\gamma_{\rm b}$;
Col. [15] $t_{\rm v}\equiv  R_{\rm blob}(1+z)/(c\mathcal{D})$.
The disk has an X-ray corona of luminosity $L_X=0.3 L_{\rm d}$.
The spectral shape of the corona is assumed to be $\propto \nu^{-1} \exp(-h\nu/150~{\rm keV})$
for the models in Supplementary Figure~\ref{F:sed-one} and Supplementary Figure~\ref{F:sed-three}, 
while it is $\propto \nu^{-0.7} \exp(-h\nu/150~{\rm keV})$
in Supplementary Figure~\ref{F:sed-two}.
}
\begin{tabular}{lllllllllllllll}
\\
\hline
State  &$R_{\rm blob}$ &$M$ &$L_{\rm d}$ ($L_{\rm d}/L_{\rm E}$) &$P^\prime_{\rm i}$  
&$B$ &$\Gamma$ &$\psi_{\rm v}$ &$\mathcal{D}$
    &$\gamma_{_0}$ &$\gamma_{\rm b}$ &$\gamma_{\rm max}$ &$s_1$  &$s_2$ &$t_{\rm v}$ \\
~[1]      &[2] &[3] &[4] &[5] &[6] &[7] &[8] &[9] &[10] &[11] &[12]  &[13] &[14] &[15]\\
\hline   
\multicolumn{5}{l}{Model 1: Supplementary Figure~\ref{F:sed-one}}\\ 
High  &0.03 &1e6  &0.13 (1)  &0.15    &3642  &10   &3 &15.7 &700  &7e3   &1e5   &0.  &2.2 &86\,s \\
Low   &0.03 &1e6  &0.13 (1)  &1.2e--2 &3642  &10   &3 &15.7 &600  &800   &900   &0   &4.2 &86\,s \\
Large &27   &1e6  &0.13 (1)  &1e--6   &2.1   &19   &2 &26.3 &40   &200   &400   &2.5 &3.5 &12.8 h\\
\hline
\multicolumn{5}{l}{Model 2: Supplementary Figure~\ref{F:sed-two}}\\ 
High  &0.048 &1e7 &45 (30)   &0.048   &6014  &12   &2 &20.4 &500  &9e3   &4e4   &0   &2.4 &106\,s \\
Low   &0.048 &1e7 &0.6 (0.4) &2.7e--3 &694   &12   &2 &20.4 &1e3  &1e3   &2.5e3 &0   &2.8 &106\,s \\
Large &18    &1e7 &1.3 (1)   &2e--6   &2.1   &17   &2 &25.1 &10   &100   &300   &0   &2.0 &8.9 h \\
\hline
\multicolumn{5}{l}{Model 3: Supplementary Figure~\ref{F:sed-three}}\\ 
High  &0.036 &1e7 &1.3 (1)   &0.18    &60    &13   &2 &21.5 &1    &60    &80    &2.5 &3   &75\,s  \\
Low   &0.036 &1e7 &1.3 (1)   &0.06    &60    &5    &2 &9.6  &1    &1     &100   &8   &8   &210\,s  \\
\hline
\end{tabular}
\vskip 0.4 true cm
\label{T:para}
\end{table}

\begin{table} 
\centering
\caption{
\textbf{Jet power in the form of radiation, Poynting flux,
bulk motion of electrons and protons (assuming one proton
per emitting electron)}.
}

\begin{tabular}{lllllll}
\\
\hline
State   &$\log P_{\rm r}$ &$\log P_{\rm B}$ &$\log P_{\rm e}$ &$\log P_{\rm p}$ 
&$L_{0.3-10}/L_{tot}$ &$L_{15-150}/L_{tot}$ \\
\hline  
Model 1: Supplementary Figure~\ref{F:sed-one}\\ 
High     &46.12 &45.65 &41.07 &41.05 &0.216   &0.318\\
Low      &44.33 &45.65 &40.13 &40.55 &0.630   &1.20e--6\\
Large    &40.82 &45.65 &41.02 &42.43 &-- &--\\
\hline
Model 2: Supplementary Figure~\ref{F:sed-two} \\
High     &45.78 &46.65 &40.24 &40.34 &0.218   &0.354 \\
Low      &44.03 &44.78 &40.77 &40.95 &0.512   &8.5e--7 \\
Large    &41.30 &45.17 &41.31 &42.71 &-- &--  \\
\hline
Model 3: Supplementary Figure~\ref{F:sed-three}\\
High     &46.14 &42.47 &44.71 &47.74 &0.098 &0.175  \\
Low      &44.17 &41.64 &43.98 &47.19 &0.695 &0.038  \\
\hline
\end{tabular}
\label{T:powers}
\end{table}

}

\clearpage
\begin{figure}
\hskip -1 cm
\vskip -1 cm
\includegraphics[height=0.75\textheight]{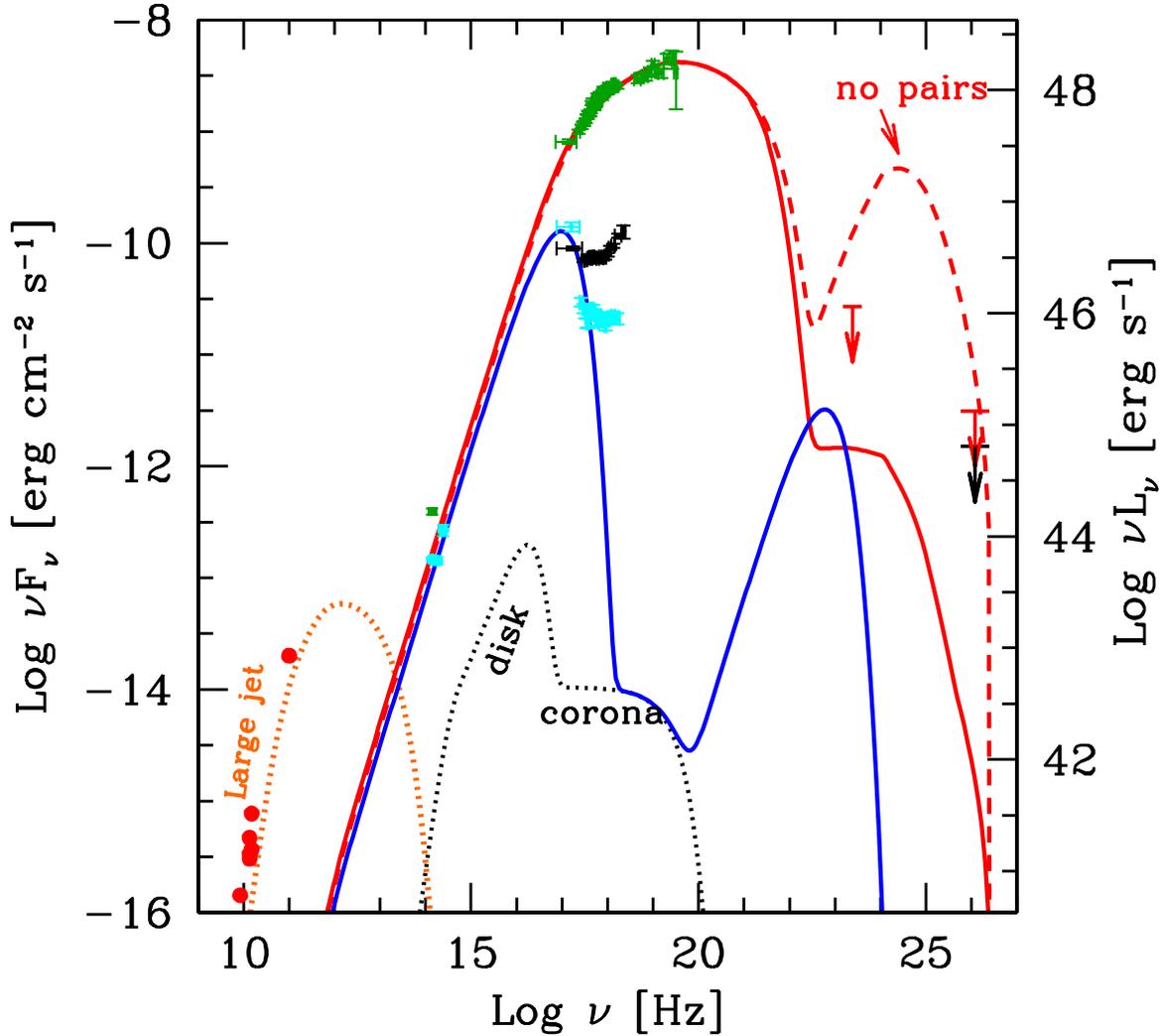}
\vskip -1. cm
\caption{\textbf{Spectral Energy Distribution for \transient.}
The green data points show the transient brightness in the near-infrared (NIR) $Ks$
band 48 hours after the trigger, while the cyan points show the $J$, $H$,
and $Ks$ band fluxes 4.5 days after trigger. The NIR flux has been dereddened with $A_V=4.5$.
In the X-ray band we show the spectrum at the peak of the bright flares
(31 hours after BAT trigger; XRT and BAT) and in the low flux state ($\sim
4.5$ days after, XRT only), 
together with the spectrum in the intermediate persistent 
flux state that began a week after the BAT trigger
(time integrated between day 6.5 and 9.5 from trigger, XRT only).
The X-ray data have been corrected for absorption with a constant $N_H=2
\times 10^{22}$~cm$^{-2}$. %
Upper limits from the {\it Fermi} LAT at $2 \times 10^{23}$~Hz and
from {\it VERITAS} at $10^{26}$~Hz\cite{Aliu11} are also shown.
The red curve shows the model discussed in the text, which
is a blazar jet model\cite{Ghisellini09} fit to our SED.
The dominant emission mechanism is synchrotron radiation peaking in
the X--ray band.  On the low frequency side, the hard slope between
the NIR and X--ray bands requires suppression of low--energy
electrons, which would otherwise overproduce the NIR flux.  On the
high frequency side, the LAT and {\it VERITAS} upper limits require that the
self--Compton component is suppressed by $\gamma$--$\gamma$ pair
production, without which the model would follow the dashed curve and
would significantly overproduce the GeV and TeV emission.  The model
includes a disk/corona component from the accretion disk (black dotted
curve), but the flux is dominated at all frequencies by the
synchrotron component from the jet.  The blue curve shows the
corresponding model in the low X--ray flux state.  The kink in the
X--ray spectrum suggests a possible additional component may be
required; it would have to be very narrow, and its origin is unclear.
The radio fluxes come from a larger region of the jet (orange dotted
line).  See Supplementary Table~\ref{T:para} and Supplementary
Table~\ref{T:powers} for the model parameters of this fit.  }
\label{F:sed-one}
\end{figure}
\begin{figure}
\hskip -1 cm
\includegraphics[height=0.75\textheight]{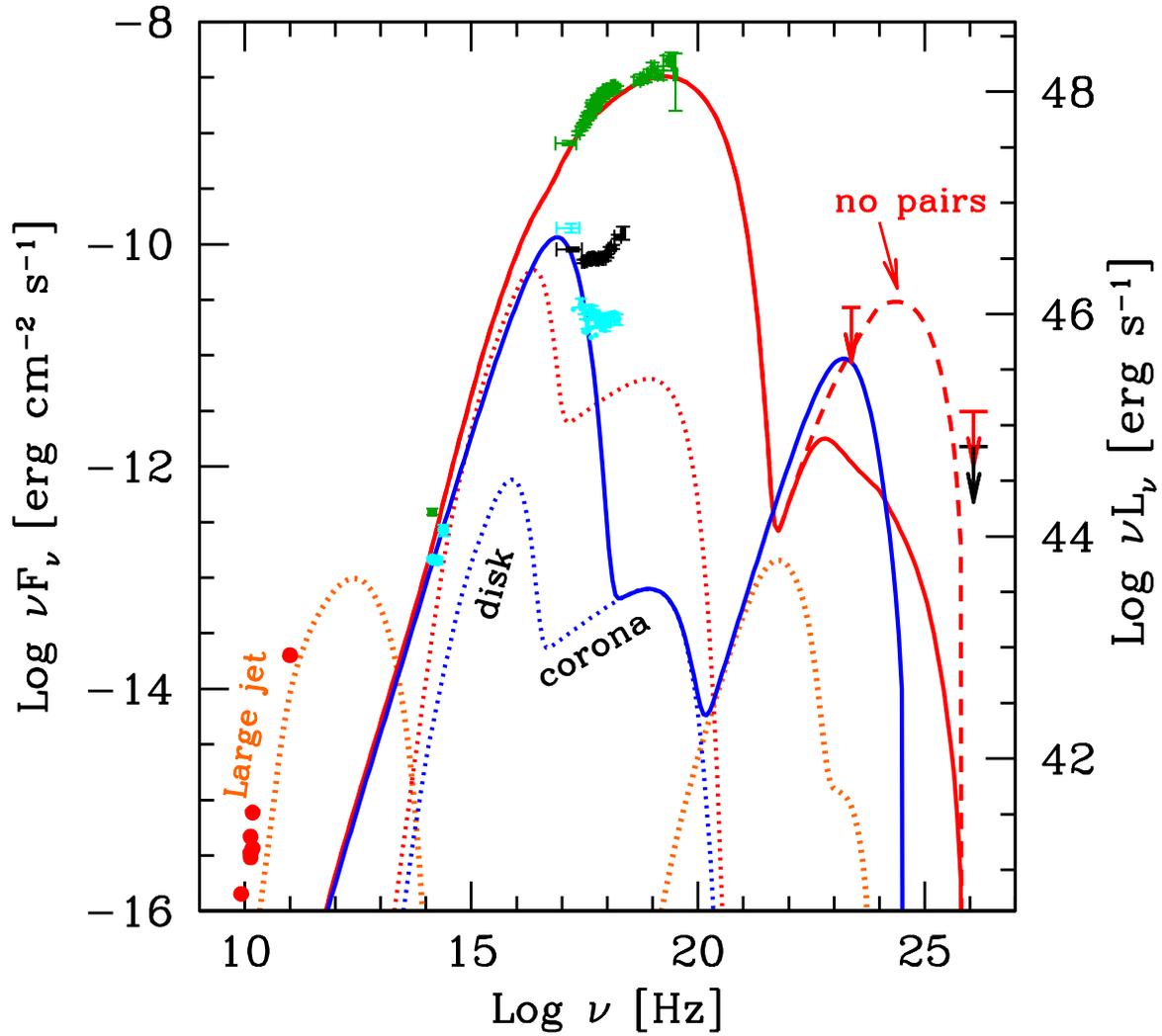}
\vskip -0.5 cm
\caption{\textbf{Data as in Supplementary Figure~\ref{F:sed-one}.}
Here the model is similar to that shown in Supplementary Figure~\ref{F:sed-one},
but the black hole mass is larger ($M_{\rm bh}= 10^7\;{\rm M}_\odot$),
and the jet luminosity is assumed to approximately track
a changing disk luminosity, from 30$\times$Eddington in the high
  state to 0.6$\times$Eddington in the low state
  (dotted lines).  
  The basic features of the jet are similar to the
  previous case, but the Poynting flux is not constant, instead it tracks
  $\dot M$. }
\label{F:sed-two}
\end{figure}
\begin{figure}
\hskip -1 cm
\includegraphics[height=0.75\textheight]{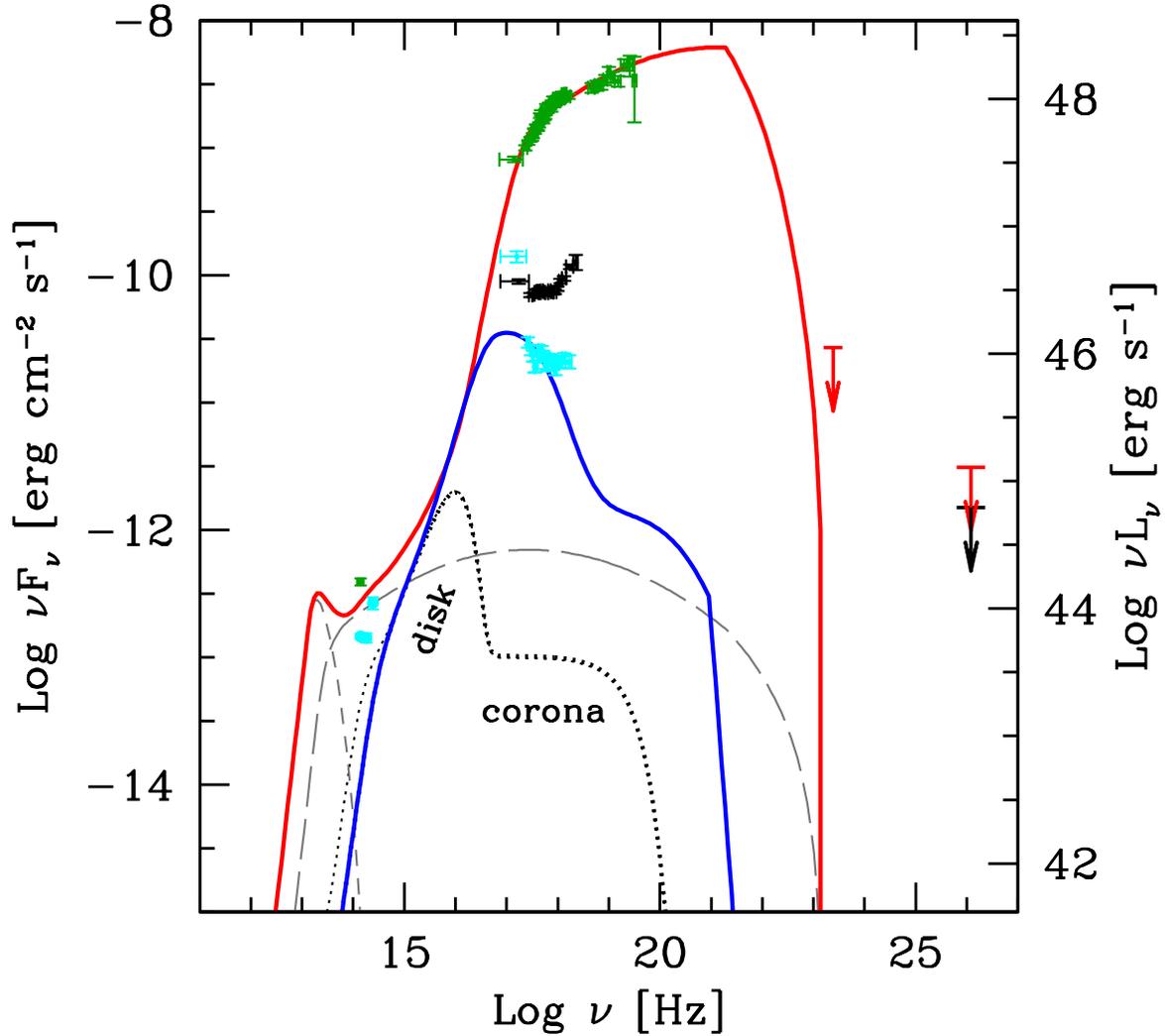}
\vskip -1 cm
\caption{\textbf{Data as in Supplementary Figure~\ref{F:sed-one}.}
In this model the NIR flux is synchrotron radiation, and the X--rays
are produced by the external Compton.
The model assumes a matter dominated jet whose dissipating region
is very close to the accretion disk (i.e. 120 $R_{\rm S}$, the black hole mass
is assumed to be $10^7 M_\odot$).
The disk itself is assumed to extend out to $\sim$500 $R_{\rm S}$.
In this case the external part of the disk produces IR photons that can
be efficiently scattered by the relativistic electrons in the jet.
In order to avoid overproducing the NIR, the magnetic field is small, and the jet must
be matter-dominated.
The high state (red line) and the low state (blue line)
differ in the amount of injected power in relativistic electrons and in the
bulk Lorentz factor.
The short-dashed gray line is the synchrotron flux in the high state,
while the long-dashed line is the corresponding SSC component.
}
\label{F:sed-three}
\end{figure}

\clearpage

\bibliography{SwiftJ1644}

\newpage